\PassOptionsToPackage{colorlinks=true, allcolors=blue}{hyperref}
\documentclass{aa}  

\usepackage{graphicx}
\usepackage{multirow}
\usepackage{subcaption}
\usepackage{bm}
\usepackage{chemformula} 
\let\ce\ch 
\usepackage{xspace}
\newcommand{\kms}{\,km\,s$^{-1}$\xspace}
\newcommand{\hrs}{\,hrs\xspace}
\newcommand{\pc}{\,pc\xspace}
\newcommand{\kpc}{\,kpc\xspace}
\newcommand{\K}{\,K\xspace}
\newcommand{\mK}{\,mK\xspace}

\newcommand{\GHz}{\,GHz\xspace}
\newcommand{\MHz}{\,MHz\xspace}
\newcommand{\kHz}{\,kHz\xspace}
\newcommand{\cm}{\,cm\xspace}
\newcommand{\mm}{\,mm\xspace}

\newcommand{\Msol}{\,M$_\odot$\xspace}
\usepackage{booktabs}
\usepackage{array}
\usepackage{upgreek}
\usepackage{orcidlink} 
\usepackage[colorlinks=true, allcolors=blue]{hyperref}
\usepackage{caption}
\usepackage{subcaption}
\usepackage{lscape}
\usepackage{longtable}
\usepackage{arydshln}
\usepackage{dashrule}
\usepackage[varg]{txfonts}

\begin{document}

\title{The Galactic Centre G+0.633-0.0604 molecular cloud: a new astrochemical gold mine}
\subtitle{I. Gas physical properties}

\author{D. San Andr{\'e}s\inst{1}\fnmsep\inst{2}\fnmsep\thanks{Corresponding author: david.sanandres@cab.inta-csic.es}\orcidlink{0000-0001-7535-4397}
        \and
        L. Colzi\inst{1}\orcidlink{0000-0001-8064-6394}
        \and
        V. M. Rivilla\inst{1}\orcidlink{0000-0002-2887-5859}
        \and
        M. Sanz-Novo\inst{3}\orcidlink{0000-0001-9629-0257}
        \and
        S. Mart{\'i}n\inst{4}\fnmsep\inst{5}\orcidlink{0000-0001-9281-2919}
        \and
        I. Jim{\'e}nez-Serra\inst{1}\orcidlink{0000-0003-4493-8714}
        \and
        \linebreak
        S. Zeng\inst{6}\orcidlink{0000-0003-3721-374X}
        }

\institute{Centro de Astrobiolog{\'i}a (CAB), CSIC-INTA, Carretera de Ajalvir km 4, 28850 Torrej{\'o}n de Ardoz, Madrid, Spain
\and
Departamento de F{\'i}sica de la Tierra y Astrof{\'i}sica, Facultad de Ciencias F{\'i}sicas, Universidad Complutense de Madrid, 28040 Madrid, Spain
\and
Center for Astrochemical Studies, Max-Planck-Institut f{\"{u}}r extraterrestrische Physik, Giessenbachstrasse 1, Garching bei Munchen, 85748, Germany 
\and
European Southern Observatory, Alonso de C{\'o}rdova, 3107, Vitacura, Santiago 763-0355, Chile
\and
Joint ALMA Observatory, Alonso de C{\'o}rdova, 3107, Vitacura, Santiago 763-0355, Chile
\and
Star and Planet Formation Laboratory, Pioneering Research Institute (PRI), RIKEN, 2-1 Hirosawa, Wako, Saitama, 351-0198, Japan
}

\date{Received 04/06/2026; Accepted 30/06/2026}

\abstract
{
In the Central Molecular Zone (CMZ) environment, shocks play a key role in both triggering star formation and providing chemical enrichment. The Sgr B2 complex is a prime template, hosting several massive protoclusters (N, M, S) and the northern G+0.693 cloud, which exhibits shock-induced prestellar signatures.
}
{
We report on G+0.633-0.0604, a newly identified shock-dominated and chemically rich cloud at the southern edge of Sgr B2, where the next star formation episodes have been proposed to occur. We characterise its physical properties and the shocks shaping it.
}
{
We analyse data from new high-sensitivity broadband spectral surveys using the Yebes 40m, IRAM 30m and APEX radio telescopes, covering a $\sim$100\GHz aggregated bandwidth across the $\sim$31$-$275\GHz range. We present analyses on \ce{CH3CCH}, \ce{CH3CN}, \ce{HC3N}, \ce{HNCO} and several isotopologues of \ce{CO} to infer kinetic temperatures and \ce{H2} densities. We complementarily used 3\mm IRAM 30m mosaics ($\sim$13\arcmin$\times$13\arcmin, $\sim$32$\times$32 pc$^2$) of Sgr B2 in \ce{HC3N}, \ce{HNCO} and \ce{C2H5OH} to probe G+0.633 environment.
}
{
We identify three velocity components towards G+0.633: a narrow main one (C1, $v_{\text{LSR}}$$\sim$48.5\kms; $\text{FWHM}$$\sim$10\kms), and two broader, much fainter components at higher velocities, C2 ($\sim$61\kms; $\sim$13\kms) and C3 ($\sim$89\kms; $\sim$18\kms), all showing similar properties ($T_{\text{kin}}$$\sim$55$-$90\K, $N_{\ce{H2}}$$\sim$(3$-$7)$\times10^{22} \; \text{cm}^{-2}$, $n_{\ce{H2}}$$\sim$(0.5$-$2.5)$\times10^{4} \; \text{cm}^{-3}$) and extended distributions. C1 delineates G+0.633 physically, and corresponds to a peak in \ce{HNCO}. This supports a shock-driven origin, likely rooted in the cloud–cloud collision event shaping Sgr B2 and also traced by C2, which mainly extends north towards G+0.693. C3 is kinematically unlinked and related to large-scale CMZ dynamics. Of the three, C1 may be consistent with a very early protocluster phase, yet to be confirmed.
}
{
G+0.633 emerges as a new shock-dominated CMZ cloud with physical resemblance to G+0.693, providing another unique laboratory to investigate how shocks drive molecular complexity and regulate the onset of cluster formation in the CMZ.
}


\authorrunning{San Andr{\'e}s et al.}\titlerunning{The physical characterisation of the Galactic Centre G+0.633-0.0604 molecular cloud}
\maketitle
\nolinenumbers

\section{Introduction}

The Galactic Centre (GC) region of the Milky Way spans its inner $\sim$300\pc around the Sgr A* central black hole, harbouring $\sim$5\% of the total molecular gas reservoir of the entire Galaxy (2$-$6$\times 10^{7}$\Msol; \citealt{Dahmen1998, Ferriere2007}) within just 0.1\% of its surface area. This gas is asymmetrically distributed in a population of several Giant Molecular Cloud complexes (with three times more mass concentrated at positive longitudes and velocities; \citealt{Henshaw2023}), which host $\sim$80\% of the dense gas in the Galaxy ($>$10$^{4} \; \text{cm}^{-3}$; \citealt{Huettemeister1993}), and give shape to the so-called Central Molecular Zone (CMZ; \citealt{Morris&Serabyn1996}). However, despite such a large reservoir of matter, star formation in the CMZ appears to be suppressed in comparison to the Galactic disc, exhibiting star formation rates at least one order of magnitude lower than expected from the measured gas mass (e.g., \citealt{Longmore2013}; \citealt{Licquia&Newman2015}; \citealt{Barnes2017}). The origin of such deficiency is still under debate, but hypothesised to be closely linked to the extreme environmental conditions that prevail in the CMZ. The high turbulence (velocity dispersions $\sim$2$-$53\kms; \citealt{Henshaw2016}) as a consequence of low-velocity shocks (e.g., \citealt{Martin-Pintado1997}; \citealt{Jones2012}), intense UV radiation (from nearby massive stellar clusters such as the Arches, Quintuplet or Young Nuclear clusters; \citealt{Lu2018}), X-ray emission (e.g., \citealt{Koyama2018}), and the enhanced cosmic-ray ionisation rates (e.g., \citealt{Goto2014, LePetit2016, Yusef-Zadeh2016, Oka2019,  Rivilla2022_PO+}), make the CMZ an extremely harsh environment for star formation and interstellar chemistry, while offering a unique window into how these processes operate under the most extreme galactic conditions.

\begin{figure*}[ht!]
    \centering
    \includegraphics[width=\textwidth]{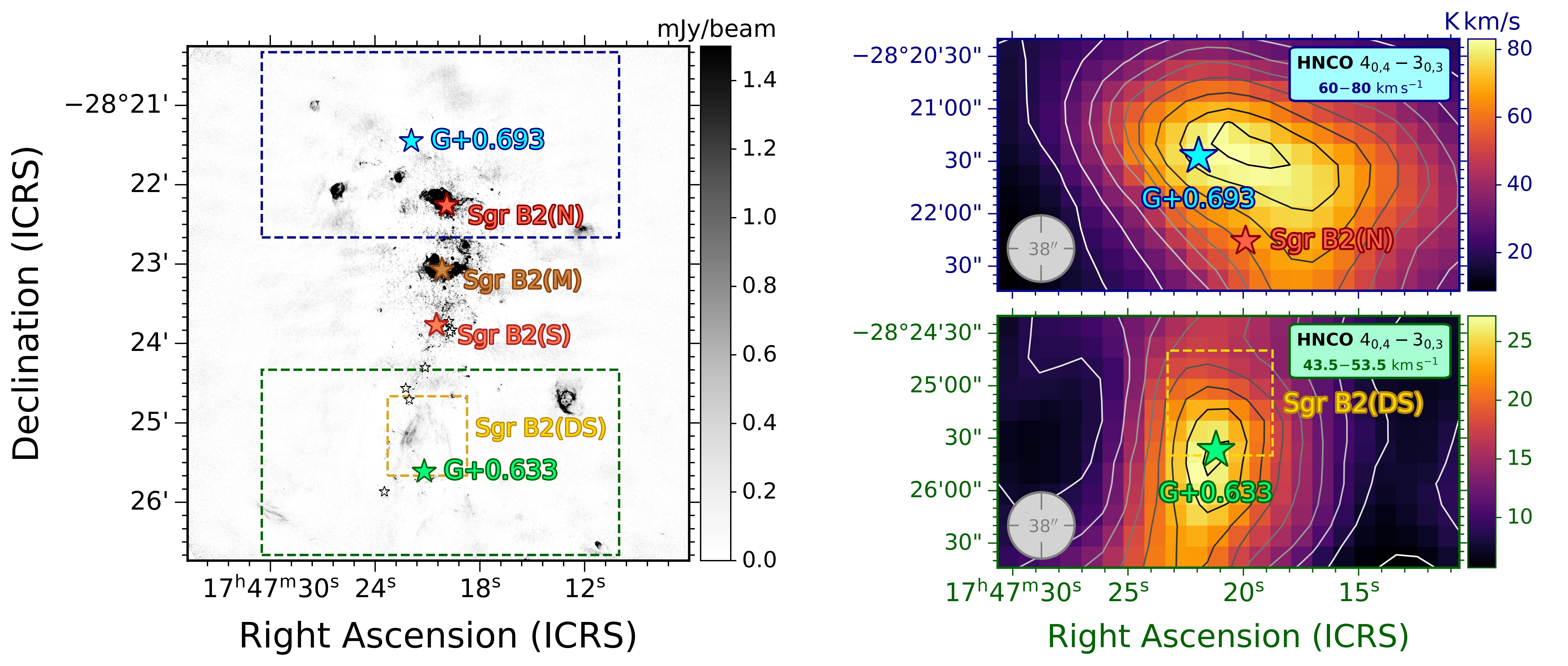}
    \caption{Left: overview of the Sgr B2 complex in ALMA 3\mm dust continuum from \citet{Ginsburg2018}. From south to north, the coloured stars indicate the positions of G+0.633, Sgr B2(S), Sgr B2(M), Sgr B2(N) and G+0.693. The white stars denote all of the dense hot cores characterised so far towards Sgr B2(S) and Sgr B2(DS) \citep{Jeff2024}. The golden dashed square represents Sgr B2(DS) extension as traced by \citet{Meng2019} VLA continuum maps at 4\GHz. The blue and green rectangles indicate the area shown in the right panels of this Figure. Right: 0th-moment map of \ce{HNCO} $4_{04}-3_{03}$ towards G+0.693 (upper panel) and G+0.633 (lower panel), computed through the MOPRA observations of the GC by \citet{Jones2012} and measured in $T_{\text{A}}^*$ units. Contours are added at 98, 90, 80, 70, 60, 50, 40, 30 and 20\% of the maximum. The beam is shown in the bottom left ($\sim$38\arcsec).
    } 
    \label{fig:SgrB2_region_overview} 
\end{figure*}

Despite the generally reduced star formation activity observed across the CMZ, this region hosts one of the most active and massive star-forming sites in the entire Galaxy: the Sgr B2 complex. Located at a distance of 8.34$\pm$0.16\kpc from the Sun and $\sim$100\pc east in projection from Sgr A* \citep{Reid2014}, Sgr B2 contains $\sim$10\% of the total CMZ dense gas. Its central $\sim$2\pc contain three high-mass protoclusters positioned along a north–south ridge in equatorial coordinates and named accordingly, Sgr B2(North), Sgr B2(Main) and Sgr B2(South), which stand out as prominent sites of ongoing star formation activity. These extensively studied massive protoclusters are surrounded by a large envelope of 15.4$'$ in diameter containing $>$99\% of this complex total mass \citep{Schmiedeke2016}, and in which some hints of star formation at very early stages, particularly towards its southernmost Sgr B2(DeepSouth) region, have recently been revealed \citep{Ginsburg2018, Meng2019, Meng2022, Jeff2024}. The morphology and kinematic features identified in Sgr B2 have been interpreted as the result of a large-scale cloud–cloud collision producing low-velocity shocks \citep{Hasegawa1994, Sato2000, Tsuboi2015, Zeng2020, Enokiya&Fukui2022, Colzi2024}, believed to have sequentially triggered such intense star formation following an ``inside-out'' pattern. Beginning at the most evolved Sgr B2(M) core \citep{deVicente2000, Schmiedeke2016}, the sequence propagates outward ``symmetrically'' \citep{Ginsburg2018, Jeff2024}: north to Sgr B2(N) and south to Sgr B2(S), eventually reaching the Sgr B2(DS) region (Fig.~\ref{fig:SgrB2_region_overview}, left).

In support of this scenario, the G+0.693-0.027 molecular cloud (hereafter G+0.693), located just $\sim$55$''$ north-east from Sgr B2(N), has become a particularly revealing case. This cloud, unlike its neighbour Sgr B2(N), does not show any signposts of ongoing star formation activity, such as ultra-compact (UC) HII regions, \ce{H2O} and Class II methanol masers, or dust continuum sources (e.g., \citealt{Ginsburg2018}; \citealt{Lu2019}). However, recent observations of dense condensations in a post-shock evolutionary stage and with high levels of deuteration \citep{Colzi2022, Colzi2024}, suggest G+0.693 may actually be the prestellar precursor of a new massive star-forming cluster within Sgr B2. In fact, G+0.693 lies right at the proposed interface of the cloud-cloud collision where the shocks are most intense \citep{Henshaw2016, Zeng2020}, evidenced by the enhanced \ce{HNCO} emission in the maps from \citealt{Jones2012} (Fig.~\ref{fig:SgrB2_region_overview}, right). Moreover, these shocks appear to have played a double role in Sgr B2, not only as triggers of its intense star formation activity, but also as drivers of its remarkable chemical wealth observed, particularly towards Sgr B2(N) (e.g., \citealt{Belloche2013, Belloche2019, Belloche2025}) and G+0.693 (e.g., \citealt{Rivilla2022_nitriles, Rivilla2023}; \citealt{Jimenez-Serra2022}; \citealt{Zeng2018, Zeng2023}; \citealt{SanAndres2023, SanAndres2024}; \citealt{Sanz-Novo2023, Sanz-Novo2024_HNSO, Sanz-Novo2025, Sanz-Novo2026}). Together, these findings establish Sgr B2 as an exceptional laboratory for exploring the interplay between shocks, star formation, and molecular complexity in the CMZ, motivating the search for additional shock-driven and chemically rich sources within its envelope.

Within this context, the Sgr B2(DS) region naturally represents the continuation of the ``inside-out'' star formation wave in Sgr B2, making it a promising location to search for a potential ``southern counterpart'' of G+0.693. Here, we report the discovery of such a source: the G+0.633-0.0604 molecular cloud (hereafter G+0.633). Located $\sim$2\arcmin\xspace south of the Sgr B2(S) protocluster, at the southern portion of Sgr B2(DS) (see Fig.~\ref{fig:SgrB2_region_overview}, left), G+0.633 occupies the next spot within the proposed sequence of shock-triggered star formation across the complex. This source was precisely selected due to its prominent \ce{HNCO} emission, the brightest within Sgr B2(DS) (Fig.~\ref{fig:SgrB2_region_overview}, right), and very narrow line profiles (similar to those of the nearby Sgr B2M (20,$-$180) source studied in \citet{Martin2008}), making it well suited for deep molecular surveys. In fact, our recent observations have revealed an exceptionally rich chemistry, with more than 120 identified molecules, placing G+0.633 among the most chemically rich sources in the CMZ (San Andrés et al., in prep.).

In this article, the first in a series devoted to G+0.633, we focus on its physical characterisation. By analysing \ce{CH3CCH}, \ce{CH3CN}, \ce{CO} isotopologues and \ce{HC3N}, we provide estimates for the gas kinetic temperature and the \ce{H2} column and volume densities. Moreover, we present an analysis on \ce{HNCO} and the spatial morphology of the molecular emission using \ce{HNCO}, \ce{HC3N} and \ce{C2H5OH} as tracers. The detailed study of the rich complex chemistry of this cloud will be presented in forthcoming works.

This article is organised as follows. In Sect.~\ref{sec:observations}, we describe the observational survey conducted towards G+0.633. In Sect.~\ref{sec:analysis_and_results} we present the analysis and results on the physical properties of this cloud, along with the molecular emission maps. In Sect.~\ref{sec:discussion} we examine these maps in detail to investigate the physical structure of G+0.633 and the underlying mechanisms driving its chemistry. Finally, in Sect.~\ref{sec:summary_and_conclusions}, we outline our conclusions.

\section{Observations and data reduction}
\label{sec:observations}

\subsection{Single-dish surveys}

We analysed data from the newly conducted ultradeep spectral line survey of G+0.633, which combines single-dish observations from the Yebes 40m (Guadalajara, Spain), IRAM\footnote{Institut de radioastronomie millimétrique} 30m (Granada, Spain) and APEX\footnote{Atacama Pathfinder EXperiment} (Chajnantor, Chile) radio telescopes. All of them were performed in position-switching mode, centred at $\alpha_{\text{ICRS}} = 17^\text{h}47^\text{m}21.18^\text{s}$, $\delta_{\text{ICRS}} = -28^\text{o}25'36.99''$. The off position ($\alpha_{\text{ICRS}}(\text{OFF}) = 17^\text{h}46^\text{m}14.92^\text{s}$, $\delta_{\text{ICRS}}(\text{OFF}) = -28^\text{o}16'36.99''$) is the same as selected in previous surveys of the neighbouring G+0.693 cloud (e.g., \citealt{Rivilla2021_Ethanolamine, Sanz-Novo2023}), devoid of emission of abundant molecules such as \ce{CS} and \ce{HC3N}. The line intensity of the spectra was directly measured in antenna temperature ($T_{\text{A}}^*$) units, since the molecular emission towards G+0.633 is extended compared to the primary beam of the three telescopes (Fig.~\ref{fig:SgrB2_region_overview} and Sect.~\ref{sec:components_discussion}).

\subsubsection{Yebes 40m observations}

We observed G+0.633 with the Yebes 40m radio telescope (project 24A010; PI: San Andr{\'e}s) for 25 different nights between February and April 2024, accounting for a total on-source observing time of $\sim$28\hrs. We used the broad-band (18.5\GHz) Nanocosmos Q-band (7\mm) HEMT receiver to cover the entire Q-band (31.07$-$50.42\GHz) in two linear polarisations \citep{Tercero2021}. The receiver was connected to 16 Fast Fourier Transform Spectrometers (FFTS) as backends (eight different with two linear polarisations each), providing a raw resolution of 38\kHz ($\sim$0.23$-$0.37\kms across the Q-band). The observations were divided into two distinct frequency setups that were usually alternated between each of the observing runs, centred at 41.4 and 42.3\GHz, to ensure unambiguous line identification against image sideband emission. The telescope pointing and focus were checked every $\sim$1$-$2\hrs through pseudo-continuum observations towards VX Sgr, a red hypergiant star close to G+0.633. The half power beam width (HPBW) of the Yebes 40m telescope varies from $\sim$55$''$ at 31\GHz down to $\sim$35$''$ at 50\GHz.

To perform the reduction of the Yebes 40m data, we exported the raw spectra from the \textsc{CLASS} module within the \textsc{GILDAS}\footnote{\url{http://www.iram.fr/IRAMFR/GILDAS}} package directly into \textsc{MADCUBA}\footnote{Madrid Data Cube Analysis (MADCUBA) on ImageJ is a software developed at the Centro de Astrobiología (CAB) in Madrid: \url{https://cab.inta-csic.es/madcuba/}} \citep{Martin2019}, where we carried out the entire processing. In a first step, we fitted and subtracted separate global baselines into each day datasets by masking the most intense lines. Then, we combined and averaged (noise-weighted) the different spectral scans along all of the observing days, obtaining a single spectrum per spectral setup, backend, and polarisation. As \citet{Tercero2021} already stated, the signal from the Nanocosmos receiver shows a combination of ripples with distinct periods that produce a wavy-like continuum, originating from internal reflections. The Yebes 40m data of G+0.633 are also affected by such ripples. To correct for them, we eliminated their associated frequency components in the Fourier space, applying, separately for each of the resulting spectra in the previous step (thereby sweeping the entire $\sim$20\GHz range four times), a large set of Fast Fourier Transformations (FFTs) over $\sim$50\MHz spectral windows. To avoid accidentally removing any actual line when applying each FFT, we simultaneously checking the consistency of the spectrum for that particular individual FFT window with the other three frequency-overlapping spectra. If coincident lines appeared, we masked them before its application. In the last step, we combined and averaged (noise-weighted) the resulting four corrected spectra after the FFT treatment for each backend, comparing beforehand the two frequency setups to remove possible spurious signals or any technical artefact. We note that we trimmed a $\sim$80\MHz portion in most of the raw spectra at $\sim$39.215\GHz, since this range is severely affected by a military radar signal near the observatory.

We smoothed the final spectra to a resolution of 256\kHz ($\sim$1.54$-$2.56\kms in the observed range), which is enough to properly resolve the observed linewidths of $\gtrsim$10\kms (see Sect.~\ref{sec:analysis_and_results}). We achieved exquisite rms noise levels between 0.5$-$1.3\mK across the $\sim$31$-$48\GHz range, rising to $\sim$2$-$2.6\mK in the Q-band upper end ($>$48\GHz) at this spectral resolution.

\subsubsection{IRAM 30m observations}

The single-dish IRAM 30m observations of G+0.633 were conducted as a result of projects 058-21 (PI: Rivilla) and 155-24 and 055-25 (PI: San Andr{\'e}s), aimed at covering the entire 3\mm window with exquisite sensitivity. In all projects, we used the broadband mm-wave heterodyne Eight MIxer Receiver (EMIR; \citealt{Carter2012}) in combination with the Fast Fourier Transform Spectrometer FTS200, to continuously cover the 3\mm band (namely E090 in EMIR) with a raw channel width of $\sim$195\kHz ($\sim$0.5$-$0.8\kms for E090). The observations on project 058-21 were carried out during the nights of 15, 16, and 18 September 2021, for a total time on source of $\sim$10.5\hrs. We employed two different spectral setups, centred at 83.50 and 91.45\GHz, to cover the lower part of the E090 window ($\sim$72$-$103\GHz). The observations on project 155-24 were conducted along five days during the first fortnight of April 2025 and on the 31st May, 1st and 2nd June 2025, achieving a total time on source of $\sim$10.5\hrs. In this project, we employed two different setups centred at 96.65 and 104.60\GHz (covering the $\sim$84$-$116.6\GHz range), completing the E090 frequency coverage while improving the overlapping part with 058-21 former setup. Finally, project 055-25 was observed for 20 days (15 in June 2025, 5 in November 2025) achieving a total time on source of $\sim$36.5\hrs. This observing run was used to further integrate along the entire 3\mm band adopting the same spectral setups from projects 058-21 and 155-24, to reach an homogeneous sensitivity across the whole survey. In all projects, we slightly shifted in frequency the adopted setups to inspect for possible contamination of spurious lines from the image band. In all cases, the telescope pointing and focus were checked every $\sim$1.5\hrs through pseudo-continuum observations towards the 1757-240 quasar. The HPBW of the IRAM 30m telescope ranges from $\sim$34$''$ down to $\sim$21$''$ in the E090 window. 

We reduced the IRAM 30m data on G+0.633 following an almost identical approach as for the Yebes 40m observations described above, including the FFT treatment to remove marked ripples that affected all side bands. We smoothed the resulting reduced spectra to 676\kHz, which translates into velocity resolutions of 1.74$-$2.82\kms over the covered frequencies. The final rms of the spectra varies $\sim$0.4$-$1.2\mK in the $\sim$72$-$114\GHz range, and increases up to $\sim$6.2\mK at the high end of the 3\mm band ($>$114\GHz) due to increased atmospheric opacity. Flux calibration errors among 058-21, 155-24 and 055-25 datasets remain $<$10\%, checked by comparing very bright lines.

\subsubsection{APEX observations}

We conducted the APEX observations of G+0.633 (project ID O-0109.F-9309A-2022; PI: Rivilla) in a total of 7 nights (5 between April and May 2022, 2 in October 2022), yielding a total time on source of $\sim$4.5\hrs. We used the 230\GHz-band New First Light APEX Submillimeter Heterodyne (nFLASH230\footnote{The nFLASH receiver was delivered by MPIfR Sub-mm technology division: \url{https://www.mpifr-bonn.mpg.de/5278273/nflash}}) receiver connected to two FFTS backends, providing simultaneous coverage through the 200$-$280\GHz range of two 7.9\GHz-wide sidebands separated by 8\GHz, with a raw spectral resolution of $\sim$250\kHz ($\sim$0.27$-$0.38\kms at these frequencies). We used two setups centred at 222.3 and 262.5\GHz to cover four spectral ranges: 209.92$-$218.33, 226.18$-$234.58, 249.92$-$258.83 and 266.17–275.08\GHz. Each setup was divided into two sub-setups shifted by $\sim$0.5\GHz, to check for possible contamination of spurious lines from the image sideband. The HPBW of these APEX observations is $\sim$22$-$29$''$ in the frequency windows observed.

We reduced the APEX observations following a methodology similar to that applied to the Yebes 40m and IRAM 30m data, with the absence of any FFT treatment but the inclusion of an additional preliminary step. Specifically, we removed some spectral channels at the lower and upper edges of each spectral scan that were significantly distorted with respect to the continuum profile. We smoothed the final reduced spectra to 1.0\MHz, which translates into velocity resolutions of 1.1$-$1.5\kms in the spectral range covered. 
We achieved a rms ranging from 0.7 to 1.0\mK along the observed windows at this spectral resolution (reaching a maximum of $\sim$1.5$-$2.1\mK at their edges).

\subsection{IRAM 30m large-scale mosaic of Sgr B2}

We complementarily used IRAM 30m mosaicking observations mapping a $\sim$13\arcmin$\times$13\arcmin\xspace ($\sim$32$\times$32 $\text{pc}^2$) region around the entire Sgr B2 complex, conducted under project 133-19 (PI: Rivilla). The data were acquired over three observing sessions on the 19-21 February 2020, yielding a total on-source integration time of $\sim$5\hrs. The data were recorded using the E090 EMIR receiver (3\mm band) and the FTS200 spectrometer, providing a spectral resolution of $\sim$195\kHz ($\sim$0.5$-$0.8\kms for this band). We tuned a single spectral setup in dual polarisation centred at $\sim$92\GHz, covering the 79.95-88.05 and 95.95-104.05\GHz spectral ranges. The HPBW of the IRAM 30m telescope ranges from $\sim$30$''$ down to $\sim$24$''$ along this range. The final mosaic was mapped by dividing the region into $\sim$40 individual subfields. Each subfield was observed using the on-the-fly (OTF) mapping technique, adopting the same off position as in the single-dish surveys described previously. The scanning of each subfield was performed in both vertical and horizontal directions with the subscans offset by 8$''$ ($\sim$1/3 of the IRAM 30m beam in the adopted spectral setup), following a zig-zag sweep pattern. The rms of the mosaic varies between $\sim$80 and 120\mK across the observed range. The maps shown in this work correspond to a fraction of the entire mosaic, particularly zooming into the Sgr B2 region.

We performed the data processing using the \textsc{CLASS} package of the \textsc{GILDAS} software. The original data cubes were resampled onto a grid with 7\arcsec of pixel size, which is $\sim$$1/3$ of the IRAM 30m beam size in the adopted setup. For the molecules analysed in this work, we generated sliced data cubes covering a $\sim$400\kms ($\sim$120\MHz) window centred at their corresponding rest frequencies. We applied a first-order baseline subtraction to each pixel (masking the lines) in all cropped daily datasets, later combined (noise-weighted) giving each species final products.

\section{Analysis and results}
\label{sec:analysis_and_results}

\renewcommand{\arraystretch}{1.25} 

\begin{table*}[t!]
 \caption[]{\label{tab:components_parameters_summary}
 Molecular emission parameters and gas physical properties of G+0.633 velocity components, compared with G+0.693.}
 \centering
 \begin{tabular}{ccccccc}
     \hline \hline
     Component & $v_{\text{LSR}}$ & $\text{FWHM}$ & $T_{\text{ex}}$ & $T_{\text{kin}}$ & $N_{\ce{H2}}$ & $n_{\ce{H2}}$ \\
     & $(\text{km} \, \text{s}^{-1})$ & $(\text{km} \, \text{s}^{-1})$ & $(\text{K})$ & $(\text{K})$ & $(\times 10^{22} \, \text{cm}^{-2})$ & $(\times 10^{4} \, \text{cm}^{-3})$ \\ 
     \hline 
     C1 & $\sim$48.5 & $\sim$10 & $\sim$3$-$20 & 55$-$90 & 6.0$\pm$1.2 & $\sim$1.1$-$2.5 \\
     C2 & $\sim$61.0 & $\sim$13 & $\sim$3$-$20 & 55$-$90 & 7$\pm$2 & $\sim$0.5$-$0.9 \\
     C3 & $\sim$89.0 & $\sim$18 & $\sim$3$-$20 & 55$-$90 & 2.8$\pm$0.7 & $\sim$0.6$-$1.0 \\
     \hline
     G+0.693-C1 & $\sim$69.0 & 15$-$25 & $\sim$3$-$20 & 70$-$140\tablefootmark{(a)} & 13.5$\pm$2.0\tablefootmark{(b)} & $\sim$2.0\tablefootmark{(c)} \\
     \hline
 \end{tabular}
  \tablefoot{The values of the molecular emission physical parameters ($v_{\text{LSR}}$, $\text{FWHM}$ and $T_{\text{ex}}$) are averaged from this work, \citet{Rivilla2026_benzo} and San Andrés et al. (in prep.). $N_{\ce{H2}}$, $T_{\text{kin}}$ and $n_{\ce{H2}}$ were estimated in this work, through the analysis of \ce{CO} isotopologues (Sect.~\ref{sec:H2_column_density_derivation}), \ce{CH3CN} and \ce{CH3CCH} (Sect.~\ref{sec:Tkin_derivation}) and \ce{HC3N} (Sect~\ref{sec:HC3N_analysis}), respectively. The corresponding values for the main velocity component observed towards G+0.693 (G+0.693-C1; \citealt{Colzi2024}) are also included for comparison. 
  \tablefoottext{a}{Derived by \citet{Zeng2018} using \ce{CH3CN} as tracer, and consistent with \citet{Colzi2024} results on \ce{HC3N} through a non-LTE approach.}
  \tablefoottext{b}{\citet{Martin2008} estimate from the analysis of \ce{C^{18}O}, adopting a 15\% of the value as uncertainty.}
  \tablefoottext{c}{Best estimate from \ce{HC3N} non-LTE analysis performed in \citet{Colzi2024} at $T_{\text{kin}} = 140$\K.}
  }
\end{table*}

In this work, we focus on a set of molecular tracers probing the physical conditions of the cloud. Specifically, we present the analyses on several \ce{CO} isotopologues (to derive the \ce{H2} column density, $N_{\ce{H2}}$), \ce{HC3N} (to retrieve the gas volume density, $n_{\ce{H2}}$), \ce{CH3CN} and \ce{CH3CCH} (to infer the gas kinetic temperature, $T_{\text{kin}}$) and \ce{HNCO} (a shock tracer). The spectroscopic information and details of their detected transitions are provided in Appendix~\ref{appendix:molecular_spectroscopy}.

We carried out the identification and spectral modelling for all of these species using the 27 May 2026 release of the Spectral Line Identification and Modelling (SLIM) tool within the \textsc{MADCUBA} package \citep{Martin2019}. For all molecules, SLIM collected their available spectroscopic data from the Cologne Database for Molecular Spectroscopy (CDMS, \citealt{Endres2016}) catalogue, which are based on different high-resolution rotational spectroscopic works (see Appendix~\ref{appendix:molecular_spectroscopy}), and generated a synthetic spectrum under the assumption of local thermodynamic equilibrium (LTE) conditions and accounting for line opacity effects, to be compared with the observed one. We derived the physical parameters describing all these molecules emission using the automatic fitting routine SLIM-AUTOFIT, which provides the best non-linear least-squares LTE fit to the data using the Levenberg-Marquardt algorithm. The free parameters used for the LTE model are the column density ($N$), excitation temperature ($T_{\text{ex}}$), local standard of rest velocity ($v_{\text{LSR}}$), and full width at half maximum ($\text{FWHM}$). The beam-filling factor was set to unity for all components, as they trace extended molecular emission relative to the telescope beams (see Sect.~\ref{sec:discussion}).

The spectral modelling of all these molecules indicates the presence of three distinct velocity components along the line of sight (see, e.g., Table~\ref{tab:components_parameters_summary} and Figs.~\ref{fig:CO_isotopologues}, \ref{fig:CH3CN_CH3CCH_ladders} and \ref{fig:HC3N_low-J}), rationalised in: 

\begin{enumerate}[(i)]
    \item A primary main component (C1), tracing gas with $v_{\text{LSR}}$ of $\sim$48.5\kms and $\text{FWHM}$ of $\sim$10\kms. This component is the brightest and thus the one leading the molecular identification in G+0.633. It is displayed in red in all figures.
    \item A secondary component (C2, depicted in green in all figures) of higher velocity ($\sim$61\kms), exhibiting a slightly broader $\text{FWHM}$ of $\sim$13\kms and usually appearing as a small shoulder next to C1 profile. Generally, the C2 peak is $\sim$3$-$5 times less intense than C1, making its contribution traceable only for those molecules in which C1 is detected with a signal-to-noise ratio (S/N) in peak intensity $\gtrsim$10.
    \item A much fainter third component (C3, shown in cyan in all figures) peaking at even higher velocities of $\sim$89\kms and with a broader $\text{FWHM}$ of $\sim$18\kms, suggesting a more turbulent gas. It can only be clearly discerned for the most abundant molecules exhibiting the brightest C1 component profiles, given the C1/C3 peak intensity ratios $\gtrsim$15 (C2/C3 $\gtrsim$ 5). This implies that C3 can only be clearly identified if C1 (C2) is with a S/N in peak intensity $\gtrsim$45 (15).
\end{enumerate}

The physical properties ($T_{\rm{kin}}$, $N_{\ce{H2}}$, $n_{\ce{H2}}$) of each of these three components, and the parameters tracing their associated molecular emission computed under LTE, are presented in Sect.~\ref{sec:G0633_physical_properties} and summarised in Table~\ref{tab:components_parameters_summary} and Appendix~\ref{appendix:LTE_results}, respectively.

\subsection{G+0.633 physical properties}
\label{sec:G0633_physical_properties}

\subsubsection{\ce{H2} column density ($N_{\ce{H2}}$)}
\label{sec:H2_column_density_derivation}

Molecular hydrogen (\ce{H2}) is the dominant component of interstellar gas, and hence often used as the standard reference for computing abundance ratios. However, its lack of a permanent electric dipole moment prevents its direct observation at \mm wavelengths. As a consequence, \ce{H2} column densities are commonly inferred from \ce{CO}, the second most abundant molecule in the ISM, which emits prominently in the radio domain.

\begin{figure*}[ht!]
    \centering
    \includegraphics[width=\textwidth]{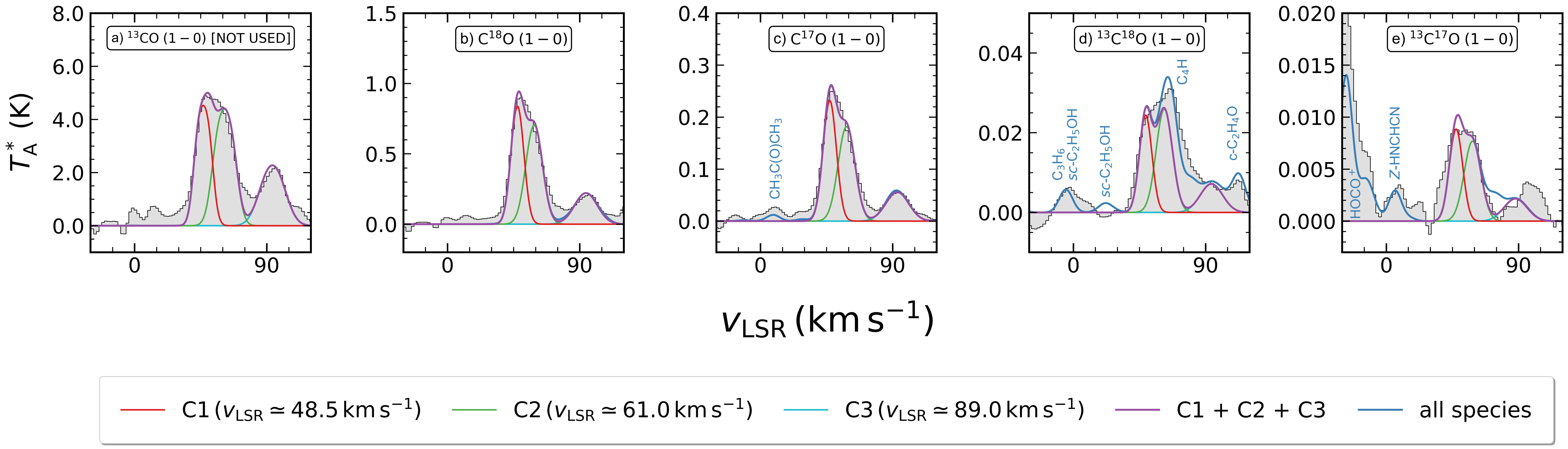}
    \caption{Transitions of \ce{CO} isotopologues detected towards the G+0.633 molecular cloud. The black histogram and grey-shaded areas delineate the observed spectrum, while the red, green and cyan solid lines outline the best LTE model obtained for the three distinct velocity components characterised in this cloud (C1, C2, and C3, respectively), together with their combined profile in purple. In blue, it is shown the contribution from the rest of species already identified towards this cloud, with their corresponding labels. Panel labels refer to the specific \ce{CO} isotopologue and transition being plotted, whose spectroscopic details are given in Table~\ref{tab:CO_isotopologues_rotational_spectroscopy}.} 
    \label{fig:CO_isotopologues} 
\end{figure*}

Nonetheless, \ce{CO} emission is heavily optically thick in G+0.633 ($\tau > 10$), thus impeding its reliable analysis. For this reason, we computed the \ce{CO} column densities (and thus $N_{\ce{H2}}$) using all detected isotopologues exhibiting optically thin emission. In this work, we analysed the lowest energy 1$-$0 transition of \ce{C^{18}O}, \ce{C^{17}O}, \ce{^{13}C^{18}O} and \ce{^{13}C^{17}O}, which is the only traced for all of them. \ce{^{13}CO}, although detected, was excluded since it is optically thick ($\tau > 1.5$). Fig.~\ref{fig:CO_isotopologues} shows the emission profiles of all these \ce{CO} isotopologues in the three distinct velocity components. We modelled the emission of each component by fixing their $\text{FWHM}$ to 10\kms for C1, 13\kms for C2 and 18\kms for C3, as commonly found across the survey (see Table~\ref{tab:components_parameters_summary}). Moreover, since only one transition is covered for each isotopologue, we fixed the $T_{\text{ex}}$ for all components to 8.8\K as found in G+0.693 for \ce{C^{18}O} \citep{Martin2008}. This assumption is well supported by the close agreement in $T_\text{ex}$ between G+0.633 and G+0.693, with similarly close values across all velocity components identified in G+0.633 (Table~\ref{tab:components_parameters_summary}; San Andrés et al., in prep.). The $v_{\text{LSR}}$ was left free for all components when performing \textsc{AUTOFIT} in most cases, with their estimates shown in Table~\ref{tab:physical_parameters_molecules_modelling}. The column densities derived for each isotopologue and velocity component are listed in Table~\ref{tab:CO_isotopologues_analysis_H2}. Subsequently, these were converted into \ce{CO} column densities adopting the GC isotopic ratios $X(\ce{^{16}O}/\ce{^{18}O}) = 250$ and $X(\ce{^{16}O}/\ce{^{17}O}) = 800$ from \citet{Wilson&Rood1994}, together with $X(\ce{^{12}C}/\ce{^{13}C}) = 40$ obtained in G+0.693 \citep{Colzi2026submitted}. For each component, we computed an average $N_{\ce{CO}}$ from all isotopologues except \ce{^{13}CO}, obtaining $N_{\ce{CO}} = (2.6 \pm 0.5) \times 10^{18} \, {\rm cm}^{-2}$, $(2.9 \pm 0.8) \times 10^{18} \, {\rm cm}^{-2}$ and $(1.2 \pm 0.3) \times 10^{18} \, {\rm cm}^{-2}$ for C1, C2 and C3, respectively. These $N_{\ce{CO}}$ estimates were finally converted into $N_{\ce{H2}}$ using a \ce{CO}-to-\ce{H2} conversion factor of $\sim$4.25$\times 10^{-5}$. This factor was calculated as $X(\ce{^{16}O}/\ce{^{18}O}) \times \ce{C^{18}O}/\ce{H2}$, adopting $\ce{C^{18}O}/\ce{H2} = 1.7\times10^{-7}$ from \citet{Frerking1982} as done in G+0.693 \citep{Martin2008}. For G+0.633, we obtained $N_{\ce{H2}} = (6.0 \pm 1.2) \times 10^{22} \, {\rm cm}^{-2}$ for C1, $N_{\ce{H2}} = (7 \pm 2) \times 10^{22} \, {\rm cm}^{-2}$ for C2 and $N_{\ce{H2}} = (2.8 \pm 0.7) \times 10^{22} \, {\rm cm}^{-2}$ for C3  (Table~\ref{tab:CO_isotopologues_analysis_H2}), which are $\sim$2.3, $\sim$1.9 and $\sim$5 times lower than in G+0.693, respectively (see Table~\ref{tab:components_parameters_summary}).

\renewcommand{\arraystretch}{1.25} 

\begin{table*}[ht!]
 \caption[]{\label{tab:CO_isotopologues_analysis_H2}Column density estimates of the different \ce{CO} isotopologues analysed in this work and used to derive $N_{\ce{H2}}$ in G+0.633.}
 \centering
 \begin{tabular}{cccccccc}
     \hline \hline
     Component & $N_{\ce{^{13}CO}}$ & $N_{\ce{C^{18}O}}$ & $N_{\ce{C^{17}O}}$ & $N_{\ce{^{13}C^{18}O}}$ & $N_{\ce{^{13}C^{17}O}}$ & $N_{\ce{CO}}$\tablefootmark{(a)} & $N_{\ce{H2}}$\tablefootmark{(b)} \\
     & $(\times 10^{16} \, \text{cm}^{-2})$ & $(\times 10^{15} \, \text{cm}^{-2})$ & $(\times 10^{15} \, \text{cm}^{-2})$ & $(\times 10^{14} \, \text{cm}^{-2})$ & $(\times 10^{13} \, \text{cm}^{-2})$ & $(\times 10^{18} \, \text{cm}^{-2})$ & $(\times 10^{22} \, \text{cm}^{-2})$ \\ 
     \hline 
     C1 & 10$\pm$1 & 9.4$\pm$0.7 & 2.5$\pm$0.1 & 2.7$\pm$0.2 & 10.0$\pm$1.1 & 2.6$\pm$0.5 & 6.0$\pm$1.2 \\
     C2 & 12$\pm$1 & 10.1$\pm$0.7 & 2.5$\pm$0.1 & 3.7$\pm$0.2 & 11.0$\pm$1.1 & 2.9$\pm$0.8 & 7$\pm$2 \\
     C3 & 5.4$\pm$0.4 & 4.2$\pm$0.6 & 1.03$\pm$0.09 & 1.4$\pm$0.2 & 4.2$\pm$1.2 & 1.2$\pm$0.3 & 2.8$\pm$0.7 \\
     \hline
 \end{tabular}
  \tablefoot{We performed the fitting of each isotopologue by fixing the $\text{FWHM}$ to 10.0\kms for C1, 13.0\kms for C2 and 18.0\kms for C3. Moreover, a $T_{\text{ex}}$ of 8.8\K was assumed for all components, attending to the value reported for \ce{C^{18}O} in G+0.693 \citep{Martin2008}, and consistent with those characterised in G+0.633 (Table~\ref{tab:components_parameters_summary}). The $v_{\text{LSR}}$ of each component was left free in most cases, with their best estimates shown in Table~\ref{tab:physical_parameters_molecules_modelling}. \tablefoottext{a}{Average \ce{CO} column density derived from all estimates obtained for the different \ce{CO} isotopologues detected in G+0.633 (excluding \ce{^{13}CO}), after applying conversion factors (isotopic ratios) of 250, 800, $10^{4}$, and $3.2\times10^{4}$ to the \ce{C^{18}O}, \ce{C^{17}O}, \ce{^{13}C^{18}O}, and \ce{^{13}C^{17}O} column densities, respectively. The uncertainties are computed as the sample standard deviation of the values derived from these four isotopologues, and are therefore dominated by systematic effects related to their dispersion and the assumed isotopic ratios.}
  \tablefoottext{b}{The $N_{\ce{H2}}$ is retrieved as $N_{\ce{CO}}/4.25$$\times10^{-5}$ (see text).}
  }
\end{table*}

\subsubsection{Kinetic temperature ($T_\text{kin}$)}
\label{sec:Tkin_derivation}

\begin{figure*}[ht!]
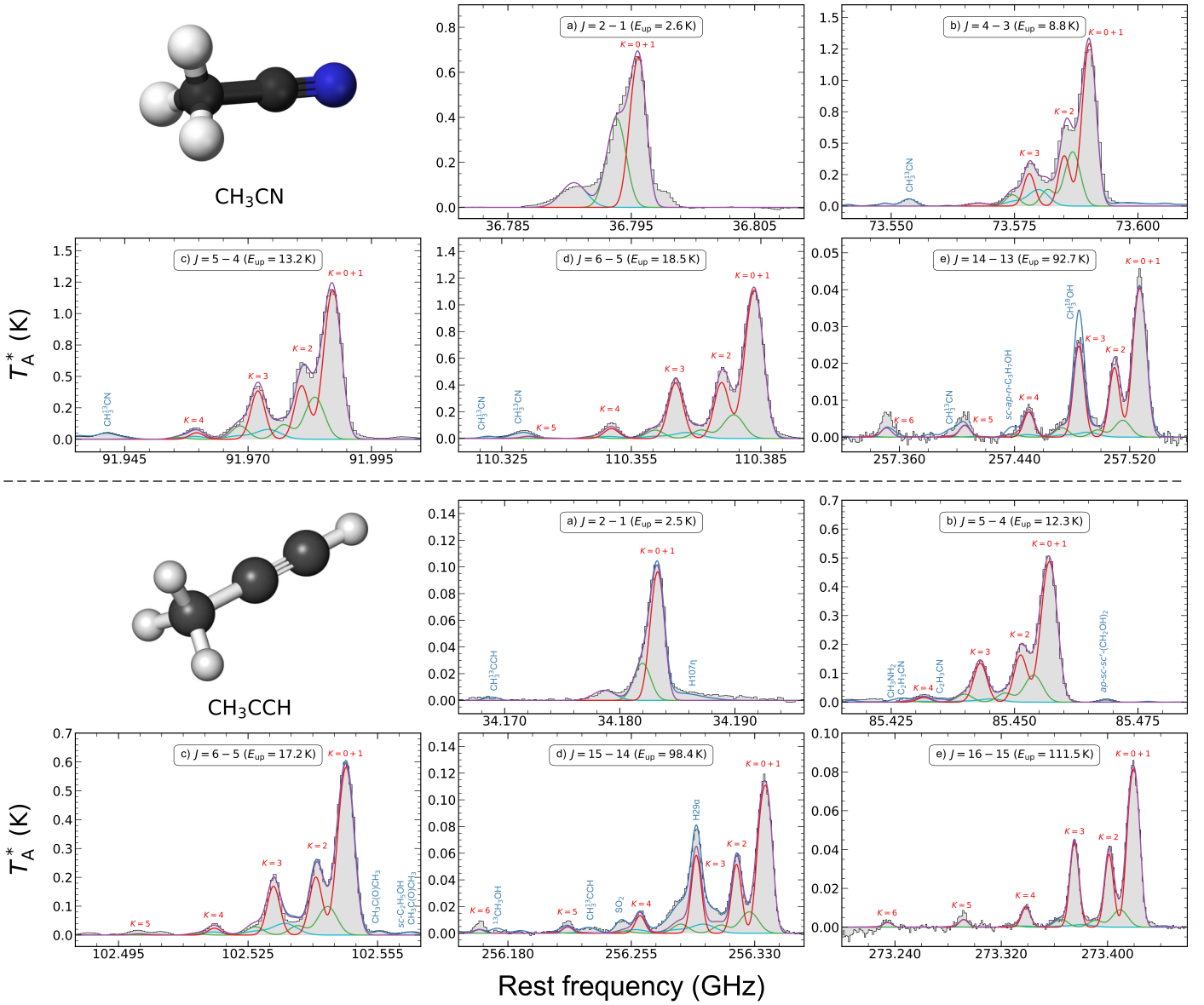

    \centering
    \begin{subfigure}[b]{\textwidth}
        \centering
        \includegraphics[width=\textwidth]{CH3CN.jpeg}
    \end{subfigure}
    \\[-1.1ex]
    \noindent\hdashrule{\textwidth}{0.5pt}{4pt 2pt}
    \\[0.9ex]
    \begin{subfigure}[b]{\textwidth}
        \centering
        \includegraphics[width=\textwidth]{CH3CCH_NOLEGEND.jpeg}
    \end{subfigure}
    \caption{$K$-ladder spectra of \ce{CH3CN} (upper panels) and \ce{CH3CCH} (lower panels) lines detected in G+0.633. The black and grey-shaded histogram, together with the red, green, cyan, purple and blue solid lines and blue labels indicate the same as in Fig.~\ref{fig:CO_isotopologues}. Panels labels delineate the main rotational transition ($J$) and the upper-level energy of its least energetic $K=0$ line ($E_{\text{up}}$). The red tags above the spectral lines identify the individual $K$ component within each $J$ transition. The molecular structures of \ce{CH3CN} and \ce{CH3CCH} are displayed in the upper-left corner of their respective panels (carbon in grey, nitrogen in blue and hydrogen in white).}
    \label{fig:CH3CN_CH3CCH_ladders}
\end{figure*}

\begin{figure*}[ht!]
    \centering
    \includegraphics[width=\textwidth]{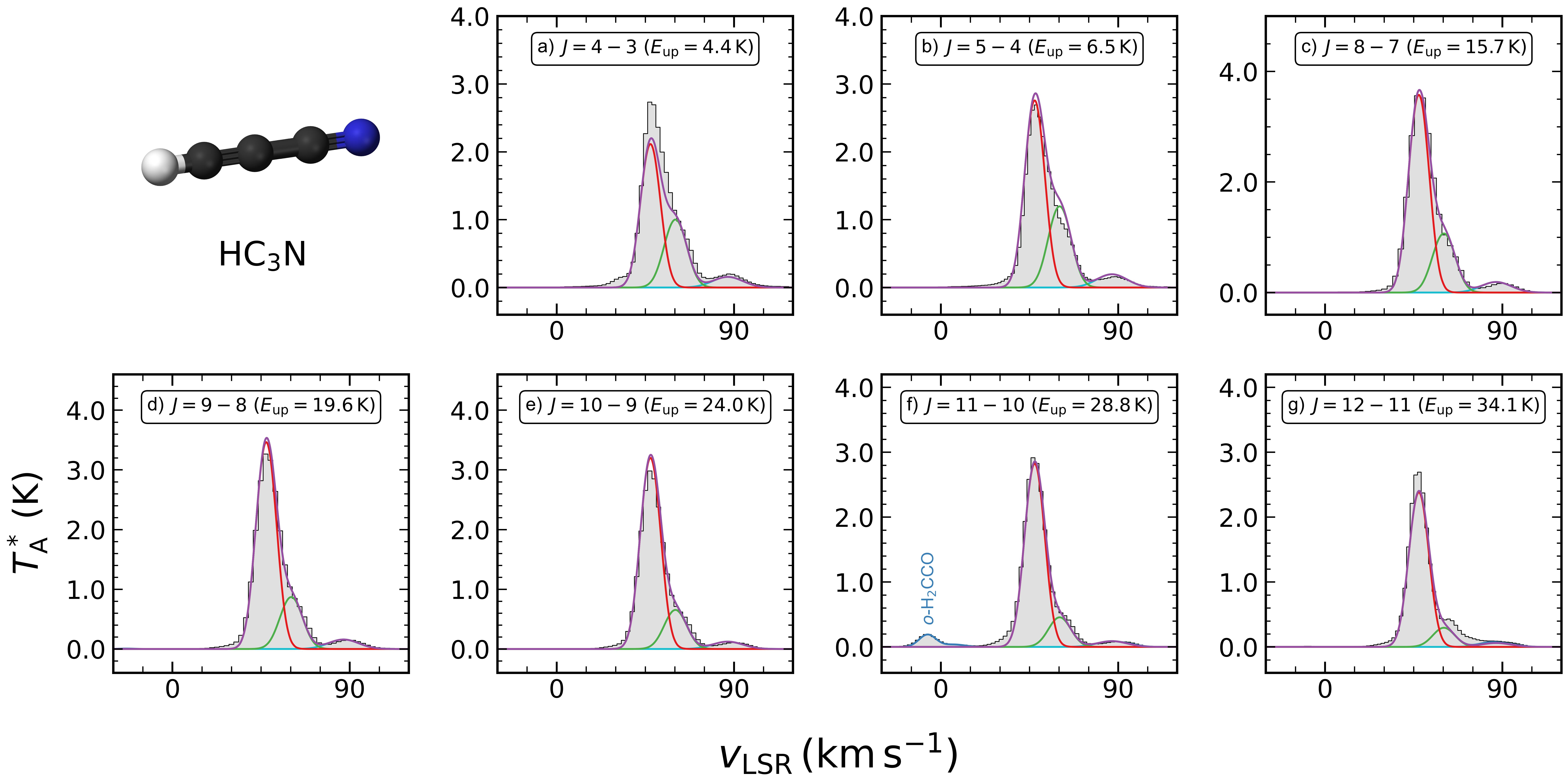}
    \caption{Same as Fig.~\ref{fig:CO_isotopologues} but for \ce{HC3N}. Panels labels indicate the specific rotational transition ($J$) being plotted and the energy of the upper level from which each transition occurs. The molecular structure of \ce{HC3N} is shown in the upper left corner.
    } 
    \label{fig:HC3N_low-J} 
\end{figure*}

\ce{CH3CN} (methyl cyanide) and \ce{CH3CCH} (propyne) are prolate symmetric top molecules that serve as excellent tracers of the kinetic temperature ($T_{\text{kin}}$) of the interstellar gas. This is because their rotational energy levels ($J$) are subdivided into successive $K$ components (the projections of the total angular momentum $J$ onto the symmetry axis) that are slightly shifted in frequency, producing a characteristic ``$K$-ladder spectrum''. Since selection rules for symmetric tops impede radiative transitions from occurring between distinct $K$-levels, their population is expected to be dominated by collisional excitation instead and thus thermalised. In this sense, the $T_{\text{ex}}$ associated with the $K$-ladder spectrum of each rotational transition of \ce{CH3CN} and \ce{CH3CCH} then provides a measurement of the $T_{\text{kin}}$ of the source (e.g., \citealt{IntroductionToAstrochemistry}).

In the current G+0.633 survey, we have detected the $K$-ladders of the $J+1 \rightarrow J$ rotational transitions with $J = 1, 3, 4, 5$ and $13$ for \ce{CH3CN}; and $J = 1, 4, 5, 14$ and $15$ for \ce{CH3CCH}. These are shown in Fig.~\ref{fig:CH3CN_CH3CCH_ladders} (upper and lower panels, respectively). We performed the fitting (adopting the three velocity component model) of each rotational transition $K$-ladder spectrum separately to mitigate possible non-LTE effects, given that the very low density found in G+0.633 makes the $T_{\text{ex}}$, defined by the same $K$ within different $J$, to be much lower compared to the $T_{\text{kin}}$, as generally found in the GC. For C1, we have identified the complete progression of $K$-levels ($K = 0, 1, ..., J$) for each $J$ transition, with the exception of the most energetic ones ($J > 13$) for which we only detected up to $K = 6$. In contrast, only the $K = 0,1,2$ and/or $K = 3$ features could be traced for the C2 and C3 components, respectively. In all cases, the $K = 0, 1$ lines are not resolved owing to the $\text{FWHM}$ ($>$10\kms). The fitting of the three velocity components was done simultaneously. However, while AUTOFIT converged for C1 in all cases when all four physical parameters were left free, the fit of C2 and C3 required the $v_{\text{LSR}}$, $\text{FWHM}$ or even the $T_{\text{ex}}$ (which traces $T_{\text{kin}}$ in this case) to be fixed in most cases. This constraint was necessary due to the significant auto-blending caused by the dominant C1 component, as evident in Fig.~\ref{fig:CH3CN_CH3CCH_ladders}. The $T_{\text{ex}}$ (i.e., $T_{\text{kin}}$) for these two components was assumed to be the same as that derived for C1, while the $v_{\text{LSR}}$ and/or $\text{FWHM}$ were fixed to the values shown in Table~\ref{tab:components_parameters_summary}, but only for those $K$-ladders for which the fit did not converge if these parameters were allowed to vary freely. The results of the individual fittings for each $K$-ladder are listed in Table~\ref{tab:physical_parameters_molecules_modelling}, yielding $T_{\text{kin}}$ estimates of 55$-$90\K for all components. 

In addition to the separate fittings presented above, we performed a complementary rotational  diagram (RD) analysis \citep{Goldsmith1999} for both \ce{CH3CN} and \ce{CH3CCH}, as implemented in \textsc{MADCUBA}, to retrieve their total column density and $T_{\text{ex}}$ (namely, $T_{\text{rot}}$). For both molecules, we computed RDs for each velocity component using just the $K = 0,1$ lines, since these are the only available for all $J+1 \rightarrow J$ transitions (see Fig.~\ref{fig:CH3CN_CH3CCH_ladders}). In all cases, we used the integrated line intensity over the linewidth as inputs, and the results are shown in Fig.~\ref{fig:CH3CN_CH3CCH_RD}. For C1, we directly performed a linear regression fit to the lines (given that all of its $K=0,1$ transitions are unblended), obtaining $T_{\text{rot}} = 13.7 \pm0.6$\K and $N = (5.7 \pm 0.9)\times 10^{13} \; \text{cm}^{-2}$ for \ce{CH3CN} and a higher $T_{\text{rot}} = 23.2 \pm0.4$\K and $N = (6.2 \pm 0.3)\times 10^{14} \; \text{cm}^{-2}$ for \ce{CH3CCH}, resulting into respective $N/N_{\ce{H2}}$ abundance ratios of $(9.5 \pm 2.4)\times10^{-10}$ and  $(1.0 \pm 0.2)\times10^{-8}$. On their side, the significant auto-blending of C1 over C2 and C3 required adopting another strategy for them. In its latest version, \textsc{MADCUBA} allows to consider the predicted emission from all molecular species previously modelled in the source and remove their contribution from the observed spectra. This generates an "unblended" spectral dataset from which a RD can be constructed for the target molecule, as successfully demonstrated in G+0.693 \citep{Rey-Montejo2024, SanAndres2024}. This also enables such an analysis for different velocity components. In this respect, we computed the RD for C2 and C3 by subtracting the contributions of i) the C1 component, and ii) the C3 or C2 component, respectively. For \ce{CH3CN}, we obtained $T_{\text{rot}} = 12.4 \pm 1.5$ and $12.0 \pm 1.2$\K; $N = (9.4 \pm 0.2)\times 10^{12}$ and $(5.45 \pm 0.35)\times 10^{12} \; \text{cm}^{-2}$ for C2 and C3, respectively. For \ce{CH3CCH}, we consistently retrieved higher $T_{\text{rot}} = 19.1 \pm 1.1$ and $19.4 \pm 0.9$\K; and $N = (1.0 \pm 0.2)\times 10^{14}$ and $(2.2 \pm 0.2)\times 10^{13} \; \text{cm}^{-2}$ for C2 and C3, respectively. The abundances relative to \ce{H2} are $(1.3 \pm 0.4)\times10^{-10}$ (C2), $(1.9 \pm 0.5)\times10^{-10}$ (C3) for \ce{CH3CN}; and $(1.4 \pm 0.5)\times10^{-9}$ (C2) and $(8 \pm 2)\times10^{-10}$ (C3) for \ce{CH3CCH}.

\subsubsection{Volume density ($n_{\ce{H2}}$)}
\label{sec:HC3N_analysis}

\ce{HC3N} (cyanoacetylene) is a linear molecule established as one of the best probes for gas densities in GC clouds (e.g., \citealt{Mills2018}) as demonstrated in G+0.693 \citep{Colzi2024}. Its diagnostic power arises from its numerous accessible rotational transitions at \cm and \mm wavelengths, evenly spaced by $\sim$9.1\GHz and connecting only adjacent energy levels given its closed-shell electronic structure. Consequently, their intensities are highly sensitive to excitation conditions, enabling the characterisation of gas over a wide range of critical densities. Besides, \ce{HC3N} collisional coefficients are accurately determined up to 300\K, suitable for GC conditions \citep{Faure2016}.

To provide an estimate for \ce{H2} volume density ($n_{\ce{H2}}$) in G+0.633, we first conducted the LTE modelling of \ce{HC3N}, depicted in Fig.~\ref{fig:HC3N_low-J}. The detected lines span a range of energy levels with upper energies of $\sim$4.4$-$34.1\K and $4 \leq J_{\text{up}} \leq 12$\footnote{Additional, much weaker, high-$J$ lines ($J_{\text{up}}=24, 25, 28, 30$) are detected in G+0.633, but were omitted as they trace gas with $\gtrsim$50 times lower column densities and some are affected by observational artifacts.}, all of which are free from emission of other molecules. As Fig.~\ref{fig:HC3N_low-J} shows, the observed \ce{HC3N} profiles also underline the three distinct velocity components identified in G+0.633. We conducted their fitting simultaneously and allowing the four physical parameters to vary freely, with the exception of the $\text{FWHM}$ for C2 and C3 which were fixed to 13 and 18\kms, respectively. The results of the fittings are summarised in Table~\ref{tab:physical_parameters_molecules_modelling}. We obtained consistent $T_{\text{ex}}$ among components of $\sim$10\K, and derived column densities of $(2.414 \pm 0.035)\times10^{14}$, $(1.11 \pm 0.04)\times10^{14}$ and $(2.22 \pm 0.34)\times10^{13} \; \text{cm}^{-2}$ for C1, C2 and C3, respectively. These are translated into respective abundances relative to \ce{H2} of $(4.0 \pm 0.8)\times10^{-9}$, $(1.6 \pm 0.5)\times10^{-9}$ and $(7.9 \pm 2.3)\times10^{-10}$.

Starting from the LTE modelling of \ce{HC3N}, we estimated the gas densities ($n_{\ce{H2}}$) of the three velocity components through a non-LTE analysis with the statistical equilibrium radiative transfer code \textsc{RADEX}\footnote{\url{https://home.strw.leidenuniv.nl/~moldata/radex.html}} \citep{vanderTak2007}, using the \ce{HC3N}$-$\ce{H2} collisional coefficients from \citet{Faure2016}. For each component, we computed a homogeneous grid of 51$\times$51 non-LTE models spanning $n_{\ce{H2}}= 10^{3}-10^{5} \; {\rm cm}^{-3}$ and $T_{\text{kin}} = 50-100 \; {\rm K}$ (in accordance with the derived $T_{\text{kin}}$ in Sect.~\ref{sec:Tkin_derivation}), adopting the corresponding \ce{HC3N} column density derived from the LTE as a fixed input. To constrain $n_{\ce{H2}}$, we then compared the predicted integrated intensities of each non-LTE model for all detected transitions shown in Fig.~\ref{fig:HC3N_low-J} with their observed values, yielding beam-averaged $n_{\ce{H2}}$ estimates representative of the spatial scales probed by the single-dish observations. The models were generated assuming a $\text{FWHM}$ of 10, 13, and 18\kms for C1, C2, and C3, respectively. The best-fitting solutions (i.e., those models that most consistently reproduce the observed emission) were identified through a fractional residual metric, described in detail in Appendix~\ref{appendix:nonLTE_HC3N}. We found that, although the non-LTE predictions show only a weak dependence on $T_{\rm kin}$, the derived densities decrease systematically with increasing temperature. Therefore, we report a range of densities corresponding to the minimum and maximum $T_\text{kin}$ inferred for each component (i.e., $\sim$55\K and $\sim$90\K). This yields $n_{\ce{H2}}$ values of $\sim$(1.1$-$2.5)$\times10^4 \; {\rm cm}^{-3}$ for C1, $\sim$(0.5$-$0.9)$\times10^4 \; {\rm cm}^{-3}$ for C2 and $\sim$(0.6$-$1)$\times10^4 \; {\rm cm}^{-3}$ for C3.

To further investigate how these density regimes are spatially distributed across the region, we generated velocity-integrated intensity maps of Sgr B2 using the $J = 11-10$ line of \ce{HC3N} for the three velocity components. These maps were computed by integrating over a velocity range centred at their respective $v_{\text{LSR}}$ ($\sim$48.5, $\sim$61.0 and $\sim$89\kms for C1, C2 and C3) and spanning $\pm1/2$ of its corresponding $\text{FWHM}$ ($\sim$10\kms for C1, $\sim$13\kms for C2 and $\sim$18\kms for C3). The resulting maps are shown in the first row of Fig.~\ref{fig:IRAM_mosaics}, and discussed in Sect.~\ref{sec:components_discussion}.

\subsection{\ce{HNCO} emission}
\label{sec:HNCO_analysis}

\ce{HNCO} (isocyanic acid) is a prolate asymmetric top molecule with a well-established role as a tracer of shock-heated regions. This species has been observed across diverse astrophysical environments, its emission being associated with other key shock tracers such as \ce{SO}, \ce{SiO} or \ce{CH3OH} (e.g., \citealt{Zinchenko2000, Meier&Turner2005}). Furthermore, \ce{HNCO} abundance is found to be particularly enhanced in the presence of large-scale low-velocity shocks ($\sim$20\kms), making it a valuable diagnostic tool to distinguish between kinematic regimes (e.g., \citealt{Kelly2017}).

\begin{figure*}[ht!]
    \centering
    \includegraphics[width=\textwidth]{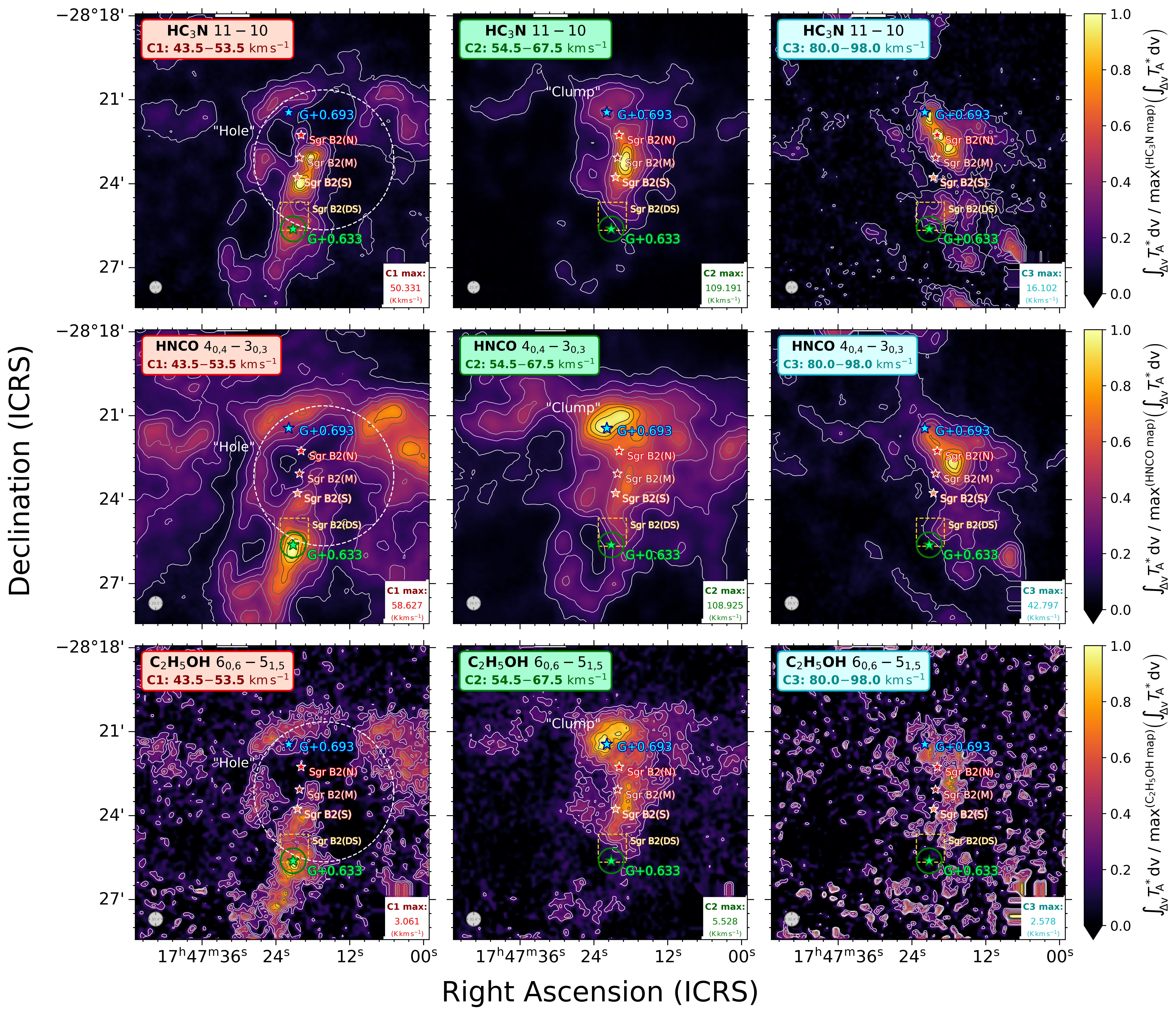}
    \caption{\ce{HC3N}, \ce{HNCO} and \ce{C2H5OH} integrated intensity maps of the Sgr B2 region obtained with the IRAM 30m radio telescope for the three velocity components identified in G+0.633 (different columns). The integration was performed over their corresponding linewidths (43.5$-$53.5\kms for C1 in red; 54.5$-$67.5\kms for C2 in green, 80.0$-$98.0\kms for C3 in cyan). Each map was normalised to its peak integrated intensity (indicated in the lower right corner), with the beam size shown in the lower left corner ($\sim$25.9\arcsec, 29.5\arcsec and 30.4\arcsec\xspace for \ce{HC3N}, \ce{HNCO} and \ce{C2H5OH} maps, respectively). Contours for \ce{HC3N} and \ce{HNCO} span 90 to 10\% of the peak in steps of 10\%, with the lowest level corresponding to $\sim$6$-$50$\sigma$ ($\sigma \equiv \text{map rms}$) depending on the molecule and component. For \ce{C2H5OH}, contours extend down to 20\% of the peak (2.9$\sigma$ for C1, 4.6$\sigma$ for C2 and 2$\sigma$ for C3). The 40$-$50\kms ``hole'' feature observed in \ce{^{13}CO} by \citet{Hasegawa1994} and \citet{Sato2000} is evident in the C1 maps and indicated by the white dashed circle, while the ``clump'' at $\sim$60$-$80\kms is visible in C2. The green circle at G+0.633 position indicates the largest beam covered in the survey ($\sim$56\arcsec, Yebes 40m beam at $\sim$31\GHz). The coloured stars and the golden dashed square are the same as in Fig.~\ref{fig:SgrB2_region_overview}.} 
    \label{fig:IRAM_mosaics} 
\end{figure*}

The G+0.633 molecular cloud is precisely located at the peak of \ce{HNCO} ($4_{0,4}-3_{0,3}$) emission in the southern part of the Sgr B2 complex, as seen in the MOPRA mosaic of the GC from \citet{Jones2012} (Fig.~\ref{fig:SgrB2_region_overview}). This can also be clearly discerned in the \ce{HNCO} velocity-integrated maps presented in Fig.~\ref{fig:IRAM_mosaics}, which were generated following the same approach as for the \ce{HC3N} 11$-$10 transition indicated above. The strong \ce{HNCO} emission and its corresponding physical parameters are also well constrained by LTE in our single-dish survey, as detailed in Appendix~\ref{appendix:HNCO_LTE_analysis}. In addition to the $K_{\text{a}} = 0$ transition of the $J = 4-3$ line targeted in the maps, we detected the two $K_{\text{a}} = 1$ transitions of the same line together with the $K_{\text{a}} = 0,1$ transitions for $J = 2-1$ and $5-4$. No $K_{\text{a}} > 1$ transition is detected in G+0.633 (nor in G+0.693). The line profiles of the $K_{\text{a}} = 0$ lines reveal the three velocity components identified in G+0.633, whereas the $K_{\text{a}} = 1$ transitions only display the C1 and C2 features (see Fig.~\ref{fig:HNCO}). We derived total \ce{HNCO} column densities of $(1.85 \pm 0.11)\times10^{15}\, \text{cm}^{-2}$, $(3.1 \pm 0.4)\times10^{14}\, \text{cm}^{-2}$ and $(6 \pm 2)\times10^{13}\, \text{cm}^{-2}$ for C1, C2 and C3, respectively. The corresponding abundances relative to \ce{H2} are $(3.1 \pm 0.7)\times10^{-8}$, $(4.4 \pm 1.6)\times10^{-9}$ and $(2.1 \pm 0.9)\times10^{-9}$. 

Overall, the prominent \ce{HNCO} emission observed towards G+0.633 is a clear indicator of the presence of intense low-velocity shocks, similarly to G+0.693 \citep{Zeng2020}. These shocks could be producing an efficient sputtering of the interstellar dust grains potentially releasing their molecular material \citep{Requena-Torres2006}, therefore being most likely responsible for the extraordinary chemical richness observed in this cloud, as will be presented in forthcoming articles. The possible origin and nature of these shocks are discussed in Sect.~\ref{sec:components_discussion}.

\section{Discussion}
\label{sec:discussion}

\subsection{Comparison between the G+0.633 and G+0.693 clouds}

The G+0.633 cloud shares several key physical characteristics with its northern counterpart, G+0.693, making them a natural pair for comparison. Beyond shocks, another remarkable feature is the presence of multiple velocity components along the line of sight. These exhibit significant differences in both clouds, providing important clues about their possible origin and nature.

In G+0.693, the dominant component (hereafter G+0.693-C1) displays broad linewidths of $\sim$18–25\kms \citep{Zeng2018, Rivilla2022_nitriles}, high $T_{\text{kin}}$ of $\sim$70$-$140\K (\citealt{Zeng2018, Colzi2024}) and relatively low $n_{\ce{H2}}$ ($\sim$2$\times10^{4} \; \text{cm}^{-3}$; \citealt{{Colzi2024}}). Combined with its extended emission, these properties point to a turbulent gas reservoir largely supported against gravitational collapse. In contrast, the secondary and tertiary components observed in this cloud (G+0.693-C2 and G+0.693-C3, respectively) exhibit much narrower line profiles ($\sim$7$-$9\kms), trace a much colder or denser gas ($T_\text{kin}$ as low as 30\K for G+0.693-C2, $n_{\ce{H2}}$ up to 4$\times10^{5} \text{cm}^{-3}$ for G+0.693-C3), and exhibit more compact distributions embedded within the extended G+0.693-C1 component \citep{Colzi2024}. Since G+0.693-C1 likely arises from the large-scale low-velocity shocks that shape Sgr B2 \citep{Zeng2020}, G+0.693-C2/C3 have been hypothesised to reflect the subsequent dissipation of turbulence, fragmentation, and/or onset of prestellar cluster formation in a post-shocked environment, given their enhanced levels of deuterium fractionation \citep{Colzi2022, Colzi2024}.

By contrast, all velocity components identified in G+0.633 display remarkably uniform physical conditions that closely resemble those of G+0.693-C1 (see Table~\ref{tab:components_parameters_summary}), thus pointing towards a similar shock-triggered nature rooted in Sgr B2 overall dynamics. In particular, all of them show relatively low \ce{H2} densities ($< 3\times10^{4} \; \text{cm}^{-3}$) and moderately high, though lower, $T_{\text{kin}}$ of $\sim$55–90\K (Sect.~\ref{sec:Tkin_derivation}), consistent with previous \ce{NH3}-based measurements across other CMZ clouds ($\sim$50–120\K; \citealt{Huettemeister1993}; \citealt{Krieger2017}). In addition, gas and dust are thermally decoupled, given the very low dust temperatures ($T_\text{dust}$) of $\sim$20\K \citep{Etxaluze2013, Battersby2025}. Along with their extended and spatially distinct morphologies (see Fig.~\ref{fig:IRAM_mosaics} and Sect.~\ref{sec:components_discussion}), these properties mainly indicate a gas reservoir unlikely to undergo imminent fragmentation or collapse. Nonetheless, the particularly narrow linewidths of G+0.633 main component (C1, $\sim$10\kms) close to those of G+0.693-C2/C3, its comparatively greater $n_{\ce{H2}}$ among all G+0.633 components (see Table~\ref{tab:components_parameters_summary}), and its more spatially confined distribution towards this cloud (Fig.~\ref{fig:IRAM_mosaics}), point to C1 tracing a significantly denser and less turbulent gas that may be more prone to gravitational collapse. Moreover, the location of G+0.633 within SgrB2(DS), where tentative signs of very early star formation have already been identified, supports a potential prestellar nature for this component. However, further analysis using established prestellar tracers, such as deuterated molecules (e.g., \citealt{Colzi2022}), in combination with high angular resolution data probing sub-pc scales, is required to clarify this point.

Collectively, these results suggest that G+0.633 may represent an even earlier evolutionary stage of shock-triggered CMZ clouds than G+0.693, while exhibiting a comparable degree of chemical enrichment. To gain deeper insight into the nature of each of its three identified components, and address their interplay in shaping G+0.633 both physically and chemically, in the next section we delve into their spatial characterisation.

\subsection{The kinematic structure of G+0.633 velocity components and the plausible origin for its chemical richness}
\label{sec:components_discussion}

To assess the physical origin of G+0.633 and the processes responsible for its chemical wealth, we examined the spatial distribution of three diagnostic molecules: \ce{HNCO}, as tracer of shock-processed gas; \ce{HC3N}, which outlines the density structure and potential gas condensations; and ethanol (\ce{C2H5OH}; here referring to the most stable $ap$ isomer, see \citealt{Rivilla2026_stereoisomers} for the updated nomenclature), as representative of the complex organic species detected in G+0.633. Since \ce{C2H5OH} mainly forms on dust grains and can be ejected into the gas phase through the action of shocks (e.g., \citealt{Jimenez-Serra2022, Agundez2023_ethanol}), its distribution provides a great proxy for the spatial extent of chemically enriched gas, and thus more complex species. The resulting velocity-integrated maps are shown in Fig.~\ref{fig:IRAM_mosaics}, and were computed for \ce{C2H5OH} mimicking the approach we followed for \ce{HC3N} and \ce{HNCO} (Sects.~\ref{sec:HC3N_analysis} and \ref{sec:HNCO_analysis}, respectively).

Inspection of Fig.~\ref{fig:IRAM_mosaics} maps reveal that among the three identified components, only C1 is directly associated with G+0.633. In fact, although \ce{HC3N}, \ce{HNCO}, and \ce{C2H5OH} display extended emission across the mapped region, it only becomes prominent towards the position of G+0.633 at C1 velocities, and peaking on it for the latter two species (first column in Fig.~\ref{fig:IRAM_mosaics}). In contrast, the spatial distributions of C2 and C3 (second and third columns in Fig.~\ref{fig:IRAM_mosaics}, respectively) are not centred on G+0.633, but rather on G+0.693 and the cores, indicating that none of these components is intrinsically related to this cloud. Their overall extended emission, unrelated to the C1 peak, results into a minimal contribution towards the position of G+0.633 as observed in all tracers, delivering the weak spectral features evident in the single-dish data. Nonetheless, the spatial continuity and soft intensity gradient in the C2 distribution from G+0.633 towards G+0.693, particularly observed in \ce{HNCO} and \ce{C2H5OH}, suggest a physical link between C1 and C2. By contrast, the very faint and spatially offset emission of C3 points towards a different origin. Therefore, C1 outlines G+0.633 physically, C2 may represent a gas interface between G+0.633 and G+0.693 (which peaks at $\sim$70\kms), whereas C3 would correspond to a separate feature.

The spatial profiles of the C1 and C2 components fit naturally within the prevailing hypothesis explaining Sgr B2 morphology and kinematics: large-scale shocks induced by a cloud–cloud collision event. The proposed scenario invokes a high-velocity ($\sim$60$-$80\kms) dense clump ploughing in the north-east direction into a more extended and less dense cloud with lower systematic velocities ($\sim$40$-$50\kms; \citealt{Armijos-Abendano2020, Zeng2020, Colzi2024}). Physical signatures supporting this scenario include a $\sim$5\arcmin\xspace ``hole'' feature at these lower velocities (delineating the penetrating region), previously identified in \ce{^{13}CO} and other tracers (e.g., \citealt{Hasegawa1994, Sato2000}), and apparent in Fig.~\ref{fig:IRAM_mosaics} for C1; and a cone-shaped structure in the position-position-velocity diagrams of Sgr B2, with its tip found at $\sim$70\kms and located close to G+0.693 (e.g. \citealt{Henshaw2016, Enokiya&Fukui2022}). 

Within this framework, the elongated morphology of C1 extending from the south towards G+0.633, together with the prominent emission lobes surrounding the ``hole'', is consistent with this component tracing the interaction interface between the two colliding clouds. In this picture, C1 mainly probes the ambient disturbed low-velocity gas as it responds to the passage of the advancing high-velocity cloud through progressive compression, stretching, and acceleration, naturally explaining the moderate density enhancement traced by \ce{HC3N}. The derived physical conditions for C1 therefore point towards an early stage of the interaction, preceding the subsequent development of denser post-shock structures. By contrast, the C2 component, characterised by a more clustered and spatially confined emission, likely traces the actual, much stronger, shock wavefronts propagating northwards to G+0.693, where the interaction appears to be strongest, 
producing the stronger gas compression inferred from \ce{HC3N} and consistent with the post-shock features identified in that cloud. The $>$80\kms gas traced by C3 is unrelated to the cloud–cloud collision event \citep{Henshaw2016, Enokiya&Fukui2022}, and arises from a more extended high-velocity structure located north-east of the Sgr B2 complex associated with the large-scale dynamics of the CMZ \citep{Sofue1995, Kruijssen2015, Kruijssen2019}.

Additional evidence supporting the cloud-cloud collision origin for G+0.633 comes from the detection of exceptionally bright Class I methanol (\ce{CH3OH}) masers (see Fig.~\ref{fig:CH3OH_ClassI_masers}). These particular masers are pumped via collisions (e.g., \citealt{Leurini2016}) and are found to be mainly excited in a wide variety of shocked environments, including molecular outflows from star-forming regions, expanding HII regions, and cloud–cloud collisions (e.g., \citealt{Voronkov2006, Sjouwerman2010}, respectively). The relative intensities of different Class I methanol maser transitions, particularly the 36 and 44\GHz lines, can also provide insight into their physical origin. In star-forming regions with protostellar outflows, the 44\GHz maser is generally stronger than the 36\GHz line (e.g., \citealt{Voronkov2014, Leurini2016}). In contrast, shocked environments not showing active star formation, such as cloud–cloud interaction zones, tend to show the opposite behaviour \citep{Sjouwerman2010, Pihlstrom2011, McEwen2016}.
In G+0.693, \citet{Zeng2020} reported that the 36\GHz methanol maser is more than an order of magnitude brighter than its 44\GHz counterpart, a strong indication of shock excitation without stellar feedback. Not surprisingly, given their expected common origin, this exact same pattern is also observed in G+0.633 (see Fig.~\ref{fig:CH3OH_ClassI_masers}). Moreover, the absence of typical high-mass star formation tracers such as \ce{OH}, \ce{H2O}, and Class II \ce{CH3OH} masers (which are radiatively pumped, \citealt{Ellingsen2005, Fontani2010}) at G+0.633 position, along with the lack of prominent dust-continuum sources (e.g., \citealt{Ginsburg2018, Lu2019, Meng2022, Jeff2024}) and Young Stellar Object (YSO) candidates \citep{DeBuizer2025}, argues against significant stellar feedback in this region. In fact, \citet{Ginsburg2018} established that high-mass star formation in the Sgr B2 complex is typically associated with $N_{\ce{H2}}\sim2\times10^{23}\;{\rm cm^{-2}}$, while no YSOs are found at $N_{\ce{H2}}<10^{23}\;{\rm cm^{-2}}$. The $N_{\ce{H2}}$ we derive for the main C1 component in G+0.633 is $\sim$2 times below this threshold (Table~\ref{tab:components_parameters_summary}), indicating that this gas is still in a stage prior to star formation.

Together, these results suggest that the shocks induced by the cloud–cloud collision are the dominant process shaping the physical structure of G+0.633 (C1 and C2) and driving its rich chemical complexity. In this framework, G+0.633 may represent a chemically enriched yet dynamically young environment where the initial stages of gas compression induced by the large-scale collision are taking place, and potentially preceding the formation of denser prestellar structures such as those observed in G+0.693. We note, however, that although C2 also appears chemically rich, its observational signatures in G+0.633 are likely limited to the most abundant molecular species.

\subsection{The potential role of Sgr B2(DS) in shaping G+0.633}
\label{sec:SgrB2(DS)_discussion}

In addition to the large-scale cloud–cloud collision likely setting the global physical conditions for G+0.633, the enhanced molecular emission observed in C1, along with the prominent Class I \ce{CH3OH} masers associated with this component, could complementarily arise from local feedback from Sgr B2(DS). In the continuum, Sgr B2(DS) appears as a bubble-like structure centred $\sim$20$''$ north of G+0.633, with an outer diameter of $\sim$1.5\pc ($\sim$40\arcsec), interpreted as an expanding HII region likely powered by a deeply embedded O7 star \citep{Meng2019}. Sgr B2(DS) exhibits a mix of compact thermal free-free and extended non-thermal synchrotron continuum emission (up to $\sim$1\arcmin, delineated by the golden dashed box in Figs.~\ref{fig:SgrB2_region_overview} and \ref{fig:IRAM_mosaics}), alongside prominent Hydrogen and Helium radio recombination lines (RRLs), indicative of highly ionised gas. These RRLs display broad profiles ($\sim$30$-$40\kms) and velocity gradients from a central $\sim$70\kms to $\sim$55\kms at the outer edge, precisely where G+0.633 is located. Consistently, numerous of these RRLs are also
detected across the G+0.633 survey (see Fig.~\ref{fig:RRLs_G0633}), exhibiting linewidths and velocities identical to those reported by \citet{Meng2019}. Interestingly, the RRLs show a peculiar intensity-frequency anti-correlation, pointing towards non-LTE excitation. This behaviour was attributed by \citet{Meng2019} to the extended synchrotron radiation observed in Sgr B2(DS), likely triggered by small-scale shocks resulting from the interaction between the expanding HII region and the denser surrounding material \citep{Padovani2019}. These shocks could then be producing a local enhancement in the molecular abundances towards Sgr B2(DS) and hence G+0.633, as seen in Fig.~\ref{fig:IRAM_mosaics}. However, we caution that the significant differences in velocity and linewidth between the RRLs and the rest of molecular lines observed in G+0.633 (also markedly differing from any of its three identified components; see Table~\ref{tab:components_parameters_summary}), prevents establishing a direct physical association with Sgr B2(DS), and the observed alignment may just result from projection effects along the line of sight. Future studies at high angular resolution will be essential to clarify the morphology and possible connection between the molecular and RRL emission observed in this region.

\section{Summary and conclusions}
\label{sec:summary_and_conclusions}

We presented a physical study on G+0.633-0.0604, a newly discovered shock-triggered and chemically rich molecular cloud at the southernmost part of the Sgr B2 complex, and which stands as the ``astrochemical twin'' of the extraordinarily rich G+0.693-0.027 cloud located north in the complex. Using high-sensitivity single-dish observations at 7, 3, and 1.3\mm from the Yebes 40m, IRAM 30m, and APEX radio telescopes, we analysed the emission of \ce{CH3CCH} and \ce{CH3CN} (to probe the kinetic temperature, $T_\text{kin}$), \ce{HC3N} and several \ce{CO} isotopologues (to constrain the gas volume, $n_{\ce{H2}}$, and column densities, $N_{\ce{H2}}$), and \ce{HNCO} (a tracer of shocks). We complemented these analyses with 3\mm mosaicking observations of the Sgr B2 region in \ce{HNCO}, \ce{HC3N} and \ce{C2H5OH}, the latter marking G+0.633 complex organic chemistry. The data reveal three distinct velocity components sharing comparable physical properties ($T_{\text{kin}}$$\sim$55$-$90\K, $n_{\ce{H2}}$$\sim$(0.5$-$2.5)$\times10^{4} \; \text{cm}^{-3}$ and $N_{\ce{H2}}$$\sim$(3$-$7)$\times10^{22} \; \text{cm}^{-2}$). The main component (C1), which displays $v_\text{LSR}$ of $\sim$48.5\kms and very narrow linewidths ($\text{FWHM}$$\sim$10\kms), dominates the emission and delineates G+0.633 physically. Its spatial correspondence with a peak in \ce{HNCO} supports a shock-driven origin, likely linked to the cloud–cloud collision event hypothesised to be shaping the Sgr B2 complex. The second, weaker component at $\sim$61\kms (C2; $\text{FWHM}$$\sim$13\kms) is also related to the same interaction, but traces the more clustered gas associated with the shock wavefronts heading north of Sgr B2. The third component (C3, $v_\text{LSR}$$\sim$89\kms, $\text{FWHM}$$\sim$18\kms) is not kinematically connected with Sgr B2, but related to the large-scale CMZ dynamics. Of the three, C1 stands out as the densest and least turbulent, with physical properties that may point to it tracing the earliest steps towards a potential prestellar phase. 

The discovery of G+0.633 as a new shock-triggered, chemically enriched cloud in an early evolutionary stage represents a cornerstone for understanding star formation in the CMZ. Its properties identify it as the southern counterpart, within Sgr B2, of the northern G+0.693 cloud, which already shows prestellar signatures. Together, these two sources provide unique laboratories to explore how shocks foster molecular complexity and set the initial conditions for cluster formation in the CMZ.

\begin{acknowledgements}

We thank the anonymous reviewer for the careful and constructive report that improved this manuscript. We are also grateful to the Yebes 40 m and IRAM 30 m telescope staff for their precious help during the different observing runs. The 40 m radio telescope at Yebes Observatory is operated by the Spanish Geographic Institute (IGN; Ministerio de Transportes, Movilidad y Agenda Urbana). IRAM is supported by INSU/CNRS (France), MPG (Germany) and IGN (Spain). D.S.A. expresses his gratitude for the financial support provided by the Comunidad de Madrid through the Grant PIPF-2022/TEC-25475. D.S.A. and V.M.R acknowledge the funds provided by the Consejo Superior de Investigaciones Cient{\'i}ficas (CSIC) and the Centro de Astrobiolog{\'i}a (CAB) through the project 20225AT015 (Proyectos intramurales especiales del CSIC). D.S.A, V.M.R and M.S.-N. are grateful for financial support provided from the grant CNS2023-144464 funded by MICIU/AEI/10.13039/501100011033 and by ``European Union NextGenerationEU/PRTR''. D.S.A, L.C, V.M.R., M.S.-N. and I.J.-S. acknowledge support from the grant PID2022-136814NB-I00 funded by the Spanish Ministry of Science, Innovation and Universities/State Agency of Research MICIU/AEI/10.13039/501100011033 and by ERDF, UE. V.M.R also acknowledges support from grant RYC2020-029387-I funded by MICIU/AEI/10.13039/501100011033 and by "ESF, Investing in your future". 
M.S.-N. acknowledges an Alexander von Humboldt postdoctoral fellowship from the Alexander von Humboldt Foundation.
I.J.-S. also acknowledges support by ERC grant OPENS, GA No. 101125858, funded by the European Union. Views and opinions expressed are however those of the author(s) only and do not necessarily reflect those of the European Union or the European Research Council Executive Agency. Neither the European Union nor the granting authority can be held responsible for them. 
S.Z. acknowledges the support by RIKEN Special Postdoctoral Researchers Programme. The project that gave rise to these results received the support of a fellowship from the ``la Caixa'' Foundation (ID 100010434). The fellowship code is LCF/BQ/PR25/12110012.

\end{acknowledgements}

\vspace{-0.52cm}

\bibliography{biblio}{}
\bibliographystyle{aa}

\begin{appendix}

\onecolumn

\clearpage

\section{Molecular spectroscopy}
\label{appendix:molecular_spectroscopy}

Table~\ref{tab:molecules_spectroscopic_info} contains the details of the spectroscopic data used for each of the molecules analysed in this work. Tables~\ref{tab:CO_isotopologues_rotational_spectroscopy}, \ref{tab:HC3N_HNCO_isotopologues_rotational_spectroscopy} and \ref{tab:CH3CN_CH3CCH_rotational_spectroscopy} indicate the spectroscopic details of the specific transitions of the \ce{CO} isotopologues, of \ce{HC3N} and \ce{HNCO}, and of \ce{CH3CN} and \ce{CH3CCH} species that have been detected in G+0.633, respectively. For each transition, we provide its rest frequency in units of $\text{GHz}$, its associated quantum numbers, the base 10 logarithm of its integrated intensity at a fixed temperature of 300\K in units of $\text{nm}^2 \, \text{MHz}$ ($\log I$), the energy in $\text{K}$ of the upper level involved in the transition ($E_{\text{up}}$), the noise measured in $\text{mK}$ within line-free spectral ranges close to it (rms) and its corresponding line opacity ($\tau$).

\setlength{\tabcolsep}{8pt} 
\renewcommand{\arraystretch}{1.3} 

\begin{table*}[ht!]
 \caption[]{\label{tab:molecules_spectroscopic_info}Spectroscopic details of the molecules studied in this work.}
 \centering
 \begin{tabular}{ccccc}
     \hline \hline
     Molecule & Catalogue & Entry & Date & Line list references \\
     & & & & \\
     \hline
     \ce{^{13}CO} & CDMS & 29501 & March 2025 & (1) \\
     \ce{C^{18}O} & CDMS & 30502 & March 2025 & (2) \\
     \ce{C^{17}O} & CDMS & 29503 & March 2025 & (3) \\
     \ce{^{13}C^{18}O} & CDMS & 31502 & April 2025 & (4) \\
     \ce{^{13}C^{17}O} & CDMS & 30503 & April 2025 & (4) \\
     \ce{HNCO} & CDMS & 43511 & May 2009 & (5), (6) \\
     \ce{HC3N} & CDMS & 51501 & October 2000 & (7)$-$(12) \\
     \ce{CH3CN} & CDMS & 41505 & November 2016 & (13)$-$(16) \\
     \ce{CH3CCH} & CDMS & 40502 & August 2008 & (17)$-$(19) \\
     \hline
 \end{tabular}
  \tablefoot{We provide the molecular catalogue, entry and corresponding date for each molecule, including the corresponding references for the computed lines list covered in our survey.}
  \tablebib{(1)~\citet{Cazzoli2004};  (2)~\citet{Cazzoli2003}; (3)~\citet{Cazzoli2002}; (4)~\citet{Puzzarini2003}; (5)~\citet{Hocking1975}; (6)~\citet{Lapinov2007}; (7)~\citet{deZafra1971}; (8)~\citet{Creswell1977}; (9)~\citet{Mallinson&deZafra1978}; (10)~\citet{Chen1991}; (11)~\citet{Yamada1995}; (12)~\citet{Thorwirth2000}; (13)~\citet{Kukolich1973}; (14)~\citet{Kukolich1982}; (15)~\citet{Boucher1977}; (16)~\citet{Cazzoli2006}; (17)~\citet{Trambarulo&Gordy1950}; (18)~\citet{Dubrulle1978}; (19)~\citet{Wlodarczak1988}.
  }
\end{table*}

\begin{table*}[ht!]
 \caption[]{\label{tab:CO_isotopologues_rotational_spectroscopy}Spectroscopic information of all \ce{CO} isotopologues transitions detected in G+0.633, and shown in Fig.~\ref{fig:CO_isotopologues}.}
 \centering
 \begin{tabular}{lcccccc}
     \hline \hline
     Isotopologue & Frequency\tablefootmark{(a)} & Transition & $\log I$ (300\K) & $E_{\text{up}}$ & rms & $\tau$ \\
     & (GHz) & ($J$$'-$$J$$''$) & ($\text{nm}^2 \, \text{MHz}$) & ($\text{K}$) & ($\text{mK}$) & ($\times 10^{-3}$) \\
     \hline
     \ce{C^{18}O} & 109.7821734(2) & $1-0$ & -5.0708 & 5.3 & 0.8 & 164(17) \\
     \ce{C^{17}O} & 112.3592848(2) & $1-0$ & -5.0420 & 5.4 & 0.9 & 19(2) \\
     \ce{^{13}C^{18}O} & 104.7113964(2) & $1-0$ & -5.1308 & 5.0 & 0.6 & 2.9(5) \\
     \ce{^{13}C^{17}O} & 107.2889455(2) & $1-0$ & -5.0996 & 5.1 & 0.7 & 0.8(3) \\
     \hline
     \multicolumn{7}{c}{Not used in the analysis} \\
     \hline
     \ce{^{13}CO} & 110.2013543(1) & $1-0$ & -5.0662 & 5.3 & 0.9 & 1730(378) \\
     \hline
 \end{tabular}
  \tablefoot{
  \tablefoottext{a}{The numbers in brackets represent the experimental uncertainty associated with the last digits, as measured by \citet{Cazzoli2004} for \ce{^{13}CO}, \citet{Cazzoli2003} for \ce{C^{18}O}, \citet{Cazzoli2002} for \ce{C^{17}O} and \citet{Puzzarini2003} for \ce{^{13}C^{18}O} and \ce{^{13}C^{17}O}.}
  }
\end{table*}

\onecolumn
\begingroup
    \centering
    \begin{longtable}{lcccccc}
        \caption{\label{tab:CH3CN_CH3CCH_rotational_spectroscopy}Spectroscopic information of \ce{CH3CN} and \ce{CH3CCH} $K$-ladders detected towards G+0.633, and shown in Fig.~\ref{fig:CH3CN_CH3CCH_ladders}.} \\
        \hline\hline
        Frequency\tablefootmark{(a)} & Transition & $\log I$ (300\K) & $E_{\text{up}}$ & rms & $\tau$ \\
        (GHz) & ($J_K$$'$$-$$J_K$$''$) & ($\text{nm}^2 \, \text{MHz}$) & ($\text{K}$) & ($\text{mK}$) & ($\times 10^{-3}$) \\
        \hline
        \endfirsthead
        \caption{continued.}\\
        \hline\hline
        Frequency\tablefootmark{(a)} & Transition & $\log I$ (300\K) & $E_{\text{up}}$ & rms & $\tau$ \\
        (GHz) & ($J_K$$'-$$J_K$$''$) & ($\text{nm}^2 \, \text{MHz}$) & ($\text{K}$) & ($\text{mK}$) & ($\times 10^{-3}$) \\
        \hline
        \endhead
        \hline
        \endfoot
        \multicolumn{6}{c}{\ce{CH3CN}} \\
        \hline
        36.7947652(1) & $2_{1}-1_{1}$ & -4.5737 & 9.8 & 0.7 & 5.0(5) \\
        36.7954748(1) & $2_{0}-1_{0}$ & -4.4384 & 2.6 & 0.7 & 7.4(5) \\
        73.5774515(1) & $4_{3}-3_{-3}$ & -3.9952 & 73.1 & 0.6 & 2.3(8) \\
        73.5774515(1) & $4_{-3}-3_{3}$ & -3.9952 & 73.1 & 0.6 & 2.3(8) \\
        73.5845434(1) & $4_{2}-3_{2}$ & -3.7094 & 37.4 & 0.6 & 7.0(1.1) \\
        73.5887996(1) & $4_{1}-3_{1}$ & -3.5814 & 16.0 & 0.6 & 12.6(1.7) \\
        73.5902186(1) & $4_{0}-3_{0}$ & -3.5429 & 8.8 & 0.6 & 15.1(1.9) \\
        91.9587263(1) & $5_{4}-4_{4}$ & -3.8674 & 127.5 & 0.5 & 0.8(4) \\
        91.9711307(1) & $5_{3}-4_{-3}$ & -3.5450 & 77.5 & 0.5 & 3.0(4) \\
        91.9711307(1) & $5_{-3}-4_{3}$ & -3.5450 & 77.5 & 0.5 & 3.0(4) \\
        91.9799946(1) & $5_{2}-4_{2}$ & -3.3752 & 41.8 & 0.5 & 6.6(5) \\
        91.9853144(1) & $5_{1}-4_{1}$ & -3.2861 & 20.4 & 0.5 & 10.4(7) \\
        91.9870879(1) & $5_{0}-4_{0}$ & -3.2580 & 13.2 & 0.5 & 12.0(8) \\
        110.3303454(1) & $6_{5}-5_{5}$ & -3.8013 & 197.1 & 0.9 & 0.24(19) \\
        110.3494707(1) & $6_{4}-5_{4}$ & -3.4485 & 132.8 & 0.9 & 1.15(19) \\
        110.3643540(1) & $6_{3}-5_{-3}$ & -3.2456 & 82.8 & 0.9 & 3.3(2) \\
        110.3643540(1) & $6_{-3}-5_{3}$ & -3.2456 & 82.8 & 0.9 & 3.3(2) \\
        110.3749894(1) & $6_{2}-5_{2}$ & -3.1201 & 47.1 & 0.9 & 6.6(2) \\
        110.3813723(1) & $6_{1}-5_{1}$ & -3.0501 & 25.7 & 0.9 & 9.9(3) \\
        110.3835002(1) & $6_{0}-5_{0}$ & -3.0275 & 18.5 & 0.9 & 11.3(3) \\
        257.3491798(2) & $14_{6}-13_{-6}$ & -2.4867 & 349.7 & 0.8 & 0.01(4) \\
        257.3491798(2) & $14_{-6}-13_{6}$ & -2.4867 & 349.7 & 0.8 & 0.01(4) \\
        257.4035848(2) & $14_{5}-13_{5}$ & -2.3440 & 271.2 & 0.8 & 0.04(4) \\
        257.4481282(2) & $14_{4}-13_{4}$ & -2.2286 & 207.0 & 0.8 & 0.08(4) \\
        257.4827919(2) & $14_{3}-13_{-3}$ & -2.1395 & 157.0 & 0.8 & 0.15(4) \\
        257.4827919(2) & $14_{-3}-13_{3}$ & -2.1395 & 157.0 & 0.8 & 0.15(4) \\
        257.5075619(2) & $14_{2}-13_{2}$ & -2.0763 & 121.3 & 0.8 & 0.23(4) \\
        257.5224279(2) & $14_{1}-13_{1}$ & -2.0384 & 99.8 & 0.8 & 0.29(4) \\
        257.5273839(2) & $14_{0}-13_{0}$ & -2.0259 & 92.7 & 0.8 & 0.32(4) \\
        \hline
        \multicolumn{6}{c}{\ce{CH3CCH}} \\
        \hline
        34.1827603(1) & $2_{1}-1_{1}$ & -6.1073 & 9.7 & 0.6 & 0.72(4) \\
        34.1834139(1) & $2_{0}-1_{0}$ & -5.9718 & 2.5 & 0.6 & 1.07(4) \\
        85.4311745(1) & $5_{4}-4_{4}$ & -5.4018 & 127.9 & 0.6 & 0.29(13) \\
        85.4426012(1) & $5_{3}-4_{3}$ & -4.7776 & 77.3 & 0.6 & 2.44(15) \\
        85.4507663(1) & $5_{2}-4_{2}$ & -4.9082 & 41.2 & 0.6 & 2.99(16) \\
        85.4556667(1) & $5_{1}-4_{1}$ & -4.8187 & 19.5 & 0.6 & 5.0(2) \\
        85.4573003(1) & $5_{0}-4_{0}$ & -4.7905 & 12.3 & 0.6 & 5.8(2) \\
        102.4990187(1) & $6_{5}-5_{5}$ & -5.3363 & 197.8 & 0.4 & 0.08(16) \\
        102.5166372(1) & $6_{4}-5_{4}$ & -4.9824 & 132.8 & 0.4 & 0.49(17) \\
        102.5303476(1) & $6_{3}-5_{3}$ & -4.4777 & 82.3 & 0.4 & 3.51(19) \\
        102.5401446(1) & $6_{2}-5_{2}$ & -4.6526 & 46.1 & 0.4 & 4.2(2) \\
        102.5460242(1) & $6_{1}-5_{1}$ & -4.5822 & 24.5 & 0.4 & 7.0(2) \\
        102.5479844(1) & $6_{0}-5_{0}$ & -4.5595 & 17.2 & 0.4 & 8.2(3) \\
        256.1607871(1) & $15_{6}-14_{6}$ & -3.6296 & 358.4 & 0.8 & 0.03(5) \\
        256.2144732(1) & $15_{5}-14_{5}$ & -3.7909 & 279.0 & 0.8 & 0.06(5) \\
        256.2584265(1) & $15_{4}-14_{4}$ & -3.6776 & 214.0 & 0.8 & 0.20(5) \\
        256.2926301(1) & $15_{3}-14_{3}$ & -3.2890 & 163.4 & 0.8 & 1.00(6) \\
        256.3170707(1) & $15_{2}-14_{2}$ & -3.5277 & 127.3 & 0.9 & 0.94(6) \\
        256.3317388(1) & $15_{1}-14_{1}$ & -3.4904 & 105.7 & 0.9 & 1.38(7) \\
        256.3366289(1) & $15_{0}-14_{0}$ & -3.4780 & 98.4 & 0.9 & 1.57(7) \\
        273.2324062(1) & $16_{6}-15_{6}$ & -3.5540 & 371.5 & 1.0 & 0.04(2) \\
        273.2896515(1) & $16_{5}-15_{5}$ & -3.7187 & 292.1 & 1.0 & 0.06(2) \\
        273.3365188(1) & $16_{4}-15_{4}$ & -3.6079 & 227.1 & 1.0 & 0.16(2) \\
        273.3729900(1) & $16_{3}-15_{3}$ & -3.2211 & 176.6 & 1.1 & 0.69(3) \\
        273.3990509(1) & $16_{2}-15_{2}$ & -3.4611 & 140.4 & 1.1 & 0.60(3) \\
        273.4146916(1) & $16_{1}-15_{1}$ & -3.4245 & 118.8 & 1.1 & 0.83(3) \\
        273.4199058(1) & $16_{0}-15_{0}$ & -3.4123 & 111.5 & 1.1 & 0.92(3) \\
        \hline
    \end{longtable}
    \tablefoot{\tablefoottext{a}{The numbers in brackets represent the experimental uncertainty associated with the last digits. The spectroscopic data were measured by \citet{Kukolich1973}, \citet{Boucher1977}, \citet{Kukolich1982} and \citet{Cazzoli2006} for \ce{CH3CN}; and by \citet{Trambarulo&Gordy1950}, \citet{Dubrulle1978} and \citet{Wlodarczak1988} for \ce{CH3CCH}.}
    }
\endgroup

\begin{table*}[h!]
    \caption[]{\label{tab:HC3N_HNCO_isotopologues_rotational_spectroscopy}Spectroscopic information of \ce{HC3N} and \ce{HNCO} lines detected in G+0.633, and shown in Figs.~\ref{fig:HC3N_low-J} and \ref{fig:HNCO}, respectively.}
    \centering
    \begin{tabular}{lccccc}
        \hline \hline
        Frequency\tablefootmark{(a)} & Transition\tablefootmark{(b)} & $\log I$ (300\K) & $E_{\text{up}}$ & rms & $\tau$ \\
        (GHz) & ($J$$'-$$J$$''$) & ($\text{nm}^2 \, \text{MHz}$) & ($\text{K}$) & ($\text{mK}$) & ($\times 10^{-3}$) \\
        \hline
        \multicolumn{6}{c}{\ce{HC3N}} \\
        \hline
        36.39232(1) & $4-3$ & -3.4520 & 4.4 & 0.7 & 212(13) \\
        45.4903138(5) & $5-4$ & -3.1642 & 6.5 & 1.2 & 287(14) \\
        72.783822(15) & $8-7$ & -2.5641 & 15.7 & 0.6 & 399(16) \\
        81.8814677(9) & $9-8$ & -2.4161 & 19.6 & 0.8 & 387(16) \\
        90.97902(2) & $10-9$ & -2.2848 & 24.0 & 0.6 & 355(16) \\
        100.076392(15) & $11-10$ & -2.1673 & 28.8 & 0.5 & 309(15) \\
        109.17363(1) & $12-11$ & -2.0612 & 34.1 & 0.7 & 256(14) \\
        \hline
        \multicolumn{6}{c}{\ce{HNCO}} \\
        \hline
        43.7990143(1) & $2_{1,2}-1_{1,1}$ & -4.8215 & 46.4 & 1.0 & 0.51(7) \\
        43.9630395(2) & $2_{0,2}-1_{0,1}$ & -4.6258 & 3.2 & 1.0 & 54(5) \\
        44.1198756(2) & $2_{1,1}-1_{1,0}$ & -4.8152 & 46.5 & 1.0 & 0.51(7) \\
        87.597330(8) & $4_{1,4}-3_{1,3}$ & -3.8307 & 53.8 & 0.5 & 2.23(7) \\
        87.925237(8) & $4_{0,4}-3_{0,3}$ & -3.7318 & 10.5 & 0.5 & 181(13) \\
        88.239020(5) & $4_{1,3}-3_{1,2}$ & -3.8244 & 53.9 & 0.5 & 2.24(7) \\
        109.495996(6) & $5_{1,5}-4_{1,4}$ & -3.5365 & 59.0 & 0.8 & 3.24(7) \\
        109.905749(7) & $5_{0,5}-4_{0,4}$ & -3.4480 & 15.8 & 0.8 & 246(18) \\
        110.298089(5) & $5_{1,4}-4_{1,3}$ & -3.5303 & 59.2 & 0.9 & 3.26(7) \\
        \hline
 \end{tabular}
  \tablefoot{
  \tablefoottext{a}{The numbers in brackets represent the experimental uncertainty associated with the last digits, as measured by \citet{deZafra1971}, \citet{Creswell1977}, \citet{Mallinson&deZafra1978}, \citet{Chen1991}, \citet{Yamada1995} and \citet{Thorwirth2000} for \ce{HC3N}; and \citet{Hocking1975} and \citet{Lapinov2007} for \ce{HNCO}.}
  \tablefoottext{b}{\ce{HNCO} transitions are designated following the common notation for asymmetric tops ($J_{K_{\rm a},K_{\rm c}}$), where $J$ refers to the total angular momentum of the molecule while $K_{\rm a,c}$ point to its projections along the $a$ and $c$ principal axes.}
  }
\end{table*}

\clearpage

\section{Multi-velocity-component LTE analysis results on \ce{^{13}CO}, \ce{C^{18}O}, \ce{C^{17}O}, \ce{^{13}C^{18}O}, \ce{^{13}C^{17}O}, \ce{HC3N}, \ce{CH3CN}, \ce{CH3CCH} and \ce{HNCO} towards G+0.633}
\label{appendix:LTE_results}

\renewcommand{\arraystretch}{1.35}

\begingroup
    \centering
    \begin{longtable}{ccccccc}
    \caption{\label{tab:physical_parameters_molecules_modelling}Results of the multi-velocity-component LTE analysis performed on \ce{^{13}CO}, \ce{C^{18}O}, \ce{C^{17}O}, \ce{^{13}C^{18}O}, \ce{^{13}C^{17}O}, \ce{HC3N}, \ce{CH3CN}, \ce{CH3CCH} and \ce{HNCO} towards G+0.633.} \\
    \hline\hline
    Molecule & Component & $\text{FWHM}$ & $v_{\text{LSR}}$ & $T_{\text{ex}}$ & $N$ & $N/N_{\ce{H2}}$ \\
    & & $(\text{km} \, \text{s}^{-1})$ & $(\text{km} \, \text{s}^{-1})$ & $(\text{K})$ & $(\times 10^{13} \, \text{cm}^{-2})$ & $(\times 10^{-10})$ \\ 
    \hline
    \endfirsthead
    \caption{continued.}\\
    \hline\hline
    Molecule & Component & $\text{FWHM}$ & $v_{\text{LSR}}$ & $T_{\text{ex}}$ & $N$ & $N/N_{\ce{H2}}$ \\
    & & $(\text{km} \, \text{s}^{-1})$ & $(\text{km} \, \text{s}^{-1})$ & $(\text{K})$ & $(\times 10^{13} \, \text{cm}^{-2})$ & $(\times 10^{-10})$ \\ 
    \hline
    \endhead
    \hline
    \endfoot
     \hline 
     \multicolumn{7}{c}{\ce{CO} isotopologues\tablefootmark{(a)}} \\
     \hline
     \multirow[t]{3}{*}{\ce{^{13}CO}} & C1 & 10.0 & 46.8$\pm$0.4 & 8.8 & (1.0$\pm$0.1)$\times$10$^{4}$ & (1.7$\pm$0.4)$\times$10$^{4}$ \\ 
     & C2 & 13.0 & 61.2$\pm$0.5 & 8.8 & (1.2$\pm$0.1)$\times$10$^{4}$ & (1.7$\pm$0.5)$\times$10$^{4}$ \\ 
     & C3 & 18.0 & 93.8$\pm$0.7 & 8.8 & (5.4$\pm$0.4)$\times$10$^{3}$ & (1.9$\pm$0.5)$\times$10$^{4}$ \\ 
     \multirow[t]{3}{*}{\ce{C^{18}O}} & C1 & 10.0 & 47.6$\pm$0.4 & 8.8 & 940$\pm$66 & (1.57$\pm$0.33)$\times$10$^{3}$ \\
     & C2 & 13.0 & 59.0$\pm$0.6 & 8.8 & (1.01$\pm$0.07)$\times$10$^{3}$ & (1.4$\pm$0.4)$\times$10$^{3}$ \\
     & C3 & 18.0 & 94.3$\pm$1.5 & 8.8 & 420$\pm$58 & (1.5$\pm$0.4)$\times$10$^{3}$ \\
     \multirow[t]{3}{*}{\ce{C^{17}O}} & C1 & 10.0 & 47.9$\pm$0.2 & 8.8 & 250$\pm$10 & 417$\pm$85 \\
     & C2 & 13.0 & 59.4$\pm$0.4 & 8.8 & 250$\pm$10 & 357$\pm$103 \\
     & C3 & 18.0 & 93.9$\pm$0.9 & 8.8 & 103$\pm$9 & 370$\pm$97 \\
     \multirow[t]{3}{*}{\ce{^{13}C^{18}O}} & C1 & 10.0 & 49.2$\pm$0.4 & 8.8 & 27$\pm$2 & 45$\pm$10 \\
     & C2 & 13.0 & 61.9$\pm$0.5 & 8.8 & 37$\pm$2 & 53$\pm$15 \\
     & C3 & 18.0 & 94.0 & 8.8 & 14$\pm$2 & 50$\pm$14 \\
     \multirow[t]{3}{*}{\ce{^{13}C^{17}O}} & C1 & 10.0 & 48.5 & 8.8 & 10.0$\pm$1.1 & 17$\pm$4 \\
     & C2 & 13.0 & 59.7$\pm$0.9 & 8.8 & 11.0$\pm$1.4 & 16$\pm$5 \\
     & C3 & 18.0 & 89.0 & 8.8 & 4.2$\pm$1.2 & 15$\pm$6 \\
     \hline 
     \multicolumn{7}{c}{Cyanoacetylene} \\
     \hline
     \multirow[t]{3}{*}{\ce{HC3N}} & C1 & 11.4$\pm$1.2 & 47.65$\pm$0.07 & 13.9$\pm$0.2 & 24.14$\pm$0.35 & 40$\pm$8 \\
     & C2 & 13.0 & 60.33$\pm$0.25 & 8.18$\pm$0.22 & 11.1$\pm$0.4 & 16$\pm$5 \\
     & C3 & 18.0 & 86.9$\pm$1.2 & 9.0$\pm$1.1 & 2.22$\pm$0.34 & 7.9$\pm$2.3 \\
     \hline 
     \multicolumn{7}{c}{Methyl cyanide\tablefootmark{(b)}} \\
     \hline
     \multirow[t]{3}{*}{\ce{CH3CN} ($J = 2-1$)} & C1 & 10.0 & 46.7$\pm$0.2 & 70.0 & $-$ & $-$ \\
     & C2 & 13.0 & 61.0 & 70.0 & $-$ & $-$ \\
     & C3 & 18.0 & 89.0 & 70.0 & $-$ & $-$ \\
     \multirow[t]{3}{*}{\ce{CH3CN} ($J = 4-3$)} & C1 & 10.9$\pm$1.4 & 47.3$\pm$0.2 & 60$\pm$7 & $-$ & $-$ \\
     & C2 & 13.0 & 61.0 & 66$\pm$22 & $-$ & $-$ \\
     & C3 & 18.0 & 89.0 & 60.0 & $-$ & $-$ \\
     \multirow[t]{3}{*}{\ce{CH3CN} ($J = 5-4$)} & C1 & 11.3$\pm$0.8 & 46.7$\pm$0.2 & 68$\pm$4 & $-$ & $-$ \\
     & C2 & 13.0 & 58.5$\pm$1.5 & 71$\pm$11 & $-$ & $-$ \\
     & C3 & 18.0 & 89.0 & 70.0 & $-$ & $-$ \\
     \multirow[t]{3}{*}{\ce{CH3CN} ($J = 6-5$)} & C1 & 11.0$\pm$0.4 & 47.0$\pm$0.1 & 67.3$\pm$1.5 & $-$ & $-$ \\
     & C2 & 13.0 & 59.9$\pm$0.3 & 68$\pm$8 & $-$ & $-$ \\
     & C3 & 18.0 & 89.0 & 70.0 & $-$ & $-$ \\
     \multirow[t]{3}{*}{\ce{CH3CN} ($J = 14-13$)} & C1 & 10.4$\pm$0.3 & 47.3$\pm$0.2 & 87.6$\pm$2.6 & $-$ & $-$ \\
     & C2 & 13.0 & 61.0$\pm$0.8 & 90.0 & $-$ & $-$ \\
     & C3 & 18.0 & 89.0 & 90.0 & $-$ & $-$ \\
     \multirow[t]{3}{*}{\ce{CH3CN}} & C1 & 10.0 & 48.5 & 13.7$\pm$0.6 & 5.7$\pm$0.9 & 9.5$\pm$2.4 \\
     & C2 & 13.0 & 61.0 & 12.4$\pm$1.5 & 0.94$\pm$0.02 & 1.3$\pm$0.4 \\
     & C3 & 18.0 & 89.0 & 12.0$\pm$1.2 & 0.545$\pm$0.035 & 1.9$\pm$0.5 \\
     \hline 
     \multicolumn{7}{c}{Propyne\tablefootmark{(b)}} \\
     \hline
     \multirow[t]{3}{*}{\ce{CH3CCH} ($J = 2-1$)} & C1 & 10.0 & 48.7$\pm$0.1 & 70.0 & $-$ & $-$ \\
     & C2 & 13.0 & 60.1$\pm$0.4 & 70.0 & $-$ & $-$ \\
     & C3 & 18.0 & 89.0 & 70.0 & $-$ & $-$ \\
     \multirow[t]{3}{*}{\ce{CH3CCH} ($J = 5-4$)} & C1 & 11.1$\pm$0.5 & 47.9$\pm$0.1 & 59.0$\pm$1.3 & $-$ & $-$ \\
     & C2 & 13.0 & 59.0$\pm$0.6 & 67$\pm$6 & $-$ & $-$ \\
     & C3 & 18.0 & 89.0 & 60.0 & $-$ & $-$ \\
     \multirow[t]{3}{*}{\ce{CH3CCH} ($J = 6-5$)} & C1 & 10.1$\pm$0.5 & 47.9$\pm$0.1 & 54.9$\pm$1.3 & $-$ & $-$ \\
     & C2 & 13.0 & 60.3$\pm$0.4 & 56$\pm$6 & $-$ & $-$ \\
     & C3 & 18.0 & 89.0 & 55.0 & $-$ & $-$ \\
     \multirow[t]{3}{*}{\ce{CH3CCH} ($J = 15-14$)} & C1 & 8.7$\pm$0.2 & 47.3$\pm$0.1 & 66.1$\pm$1.4 & $-$ & $-$ \\
     & C2 & 13.0 & 58.3$\pm$0.6 & 65.0 & $-$ & $-$ \\
     & C3 & 18.0 & 90.8$\pm$1.1 & 65.0 & $-$ & $-$ \\
     \multirow[t]{3}{*}{\ce{CH3CCH} ($J = 16-15$)} & C1 & 8.7$\pm$0.2 & 47.2$\pm$0.1 & 69.4$\pm$1.4 & $-$ & $-$ \\
     & C2 & 13.0 & 58.4$\pm$0.8 & 70.0 & $-$ & $-$ \\
     & C3 & 18.0 & 89.0 & 70.0 & $-$ & $-$ \\
     \multirow[t]{3}{*}{\ce{CH3CCH}} & C1 & 10.0 & 48.5 & 23.2$\pm$0.4 & 62$\pm$3 & 103$\pm$21 \\
     & C2 & 13.0 & 61.0 & 19.1$\pm$1.1 & 10$\pm$2 & 14$\pm$5 \\
     & C3 & 18.0 & 89.0 & 19.4$\pm$0.9 & 2.2$\pm$0.2 & 8$\pm$2 \\
     \hline
     \multicolumn{7}{c}{Isocyanic acid\tablefootmark{(c)}} \\
     \hline
     \multirow[t]{3}{*}{\ce{HNCO} ($K_{\text{a}} = 0$)} & C1 & 11.25$\pm$0.16 & 47.9$\pm$0.1 & 34$\pm$2 & 306$\pm$16 & 510$\pm$110 \\
     & C2 & 13.0 & 60.2$\pm$0.4 & 13.1$\pm$1.2 & 44.7$\pm$2.2 & 64$\pm$19 \\
     & C3 & 18.0 & 90.7$\pm$1.9 & 5.6$\pm$1.3 & 9.6$\pm$2.3 & 34$\pm$12 \\
     \multirow[t]{3}{*}{\ce{HNCO} ($K_{\text{a}} = 1$)} & C1 & 10.13$\pm$0.13 & 47.8$\pm$0.1 & 50$\pm$8 & 9.5$\pm$1.4 & 16$\pm$4 \\
     & C2 & 13.0 & 58.8$\pm$0.5 & 47$\pm$27 & 2.0$\pm$1.1 & 2.9$\pm$1.8 \\
     \multirow[t]{3}{*}{\ce{HNCO}} & C1 & $-$ & $-$ & $-$ & 316$\pm$17 & 526$\pm$114 \\
     & C2 & $-$ & $-$ & $-$ & 47$\pm$3 & 67$\pm$21 \\
     & C3 & $-$ & $-$ & $-$ & 9.6$\pm$2.3 & 34$\pm$12 \\
     \hline
    \end{longtable}
  \tablefoot{Parameters shown without uncertainties were assumed in the LTE fit. The last column indicates the molecular abundances relative to \ce{H2}, which were computed by adopting $N_{\ce{H2}} = (6.0 \pm 1.2)\times10^{22} \; \text{cm}^{-2}$ for C1, $N_{\ce{H2}} = (7 \pm 2)\times10^{22} \; \text{cm}^{-2}$ for C2 and $N_{\ce{H2}} = (2.8 \pm 0.7)\times10^{22} \; \text{cm}^{-2}$ for C3, as derived in Sect.~\ref{sec:H2_column_density_derivation} (see also Table~\ref{tab:CO_isotopologues_analysis_H2}). 
  \tablefoottext{a}{For all \ce{CO} isotopologues, we assumed the same $T_\text{ex}$ as derived by \citet{Martin2008} from \ce{C^{18}O} analysis in G+0.693.}
  \tablefoottext{b}{\ce{CH3CN} and \ce{CH3CCH} entries for which the $J$ value is indicated gather the results from the independent analyses performed of the $K$-ladder spectrum associated with each particular $J$ level (see Sect.~\ref{sec:Tkin_derivation}). In all these separate analyses, $T_\text{ex}$ is a proxy of $T_\text{kin}$, and the resulting $N$ and $N/N_{\ce{H2}}$ values are omitted since these lack of any physical relevance. The two remaining entries without indicated $J$ summarise these species $T_\text{ex}$, total column densities and abundances relative to \ce{H2} computed from the rotational diagram analysis presented in Appendix~\ref{appendix:CH3CN_CH3CCH_RD}.}
  \tablefoottext{c}{The $K_\text{a} = 0$ and $K_\text{a} = 1$ entries for \ce{HNCO} delineate the results from the separate fittings performed on these two particular groups of lines (which include all $J$ levels), as explained in Appendix~\ref{appendix:HNCO_LTE_analysis}. The last \ce{HNCO} entry without $K_\text{a}$ delivers the total column densities and corresponding abundances relative to \ce{H2} for each component, calculated from the sum of those derived for the $K_{\text{a}} = 0$ and $K_{\text{a}} = 1$ entries displayed above.}
  }
\endgroup

\clearpage

\section{\ce{CH3CN} and \ce{CH3CCH} rotational diagram analysis}
\label{appendix:CH3CN_CH3CCH_RD}

\begin{figure*}[ht!]
    \centering
    \begin{subfigure}[b]{0.47\textwidth}
        \centering
        \includegraphics[width=\textwidth]{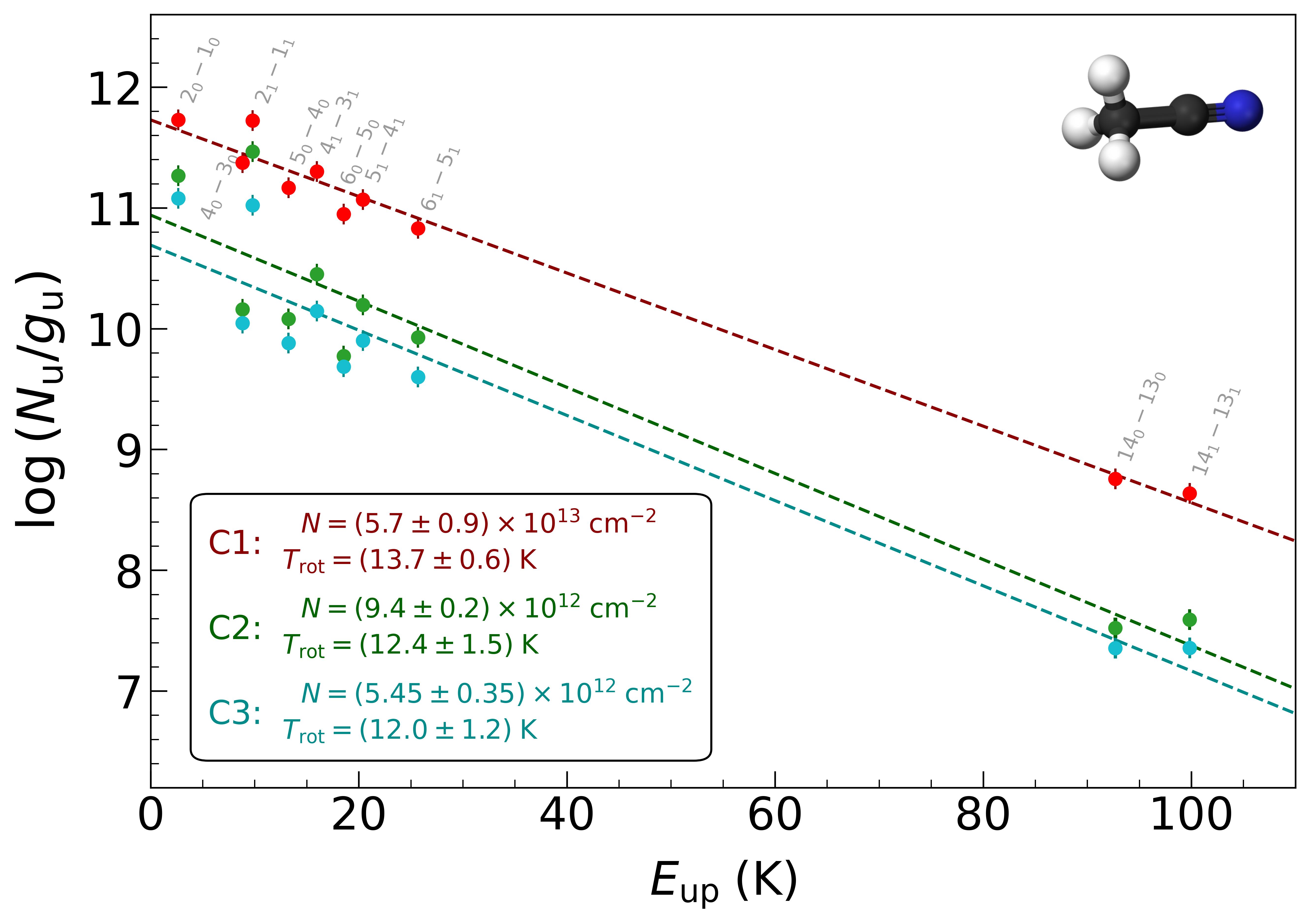}
    \end{subfigure}
    \hspace{0.8cm}
    \begin{subfigure}[b]{0.47\textwidth}
        \centering
        \includegraphics[width=\textwidth]{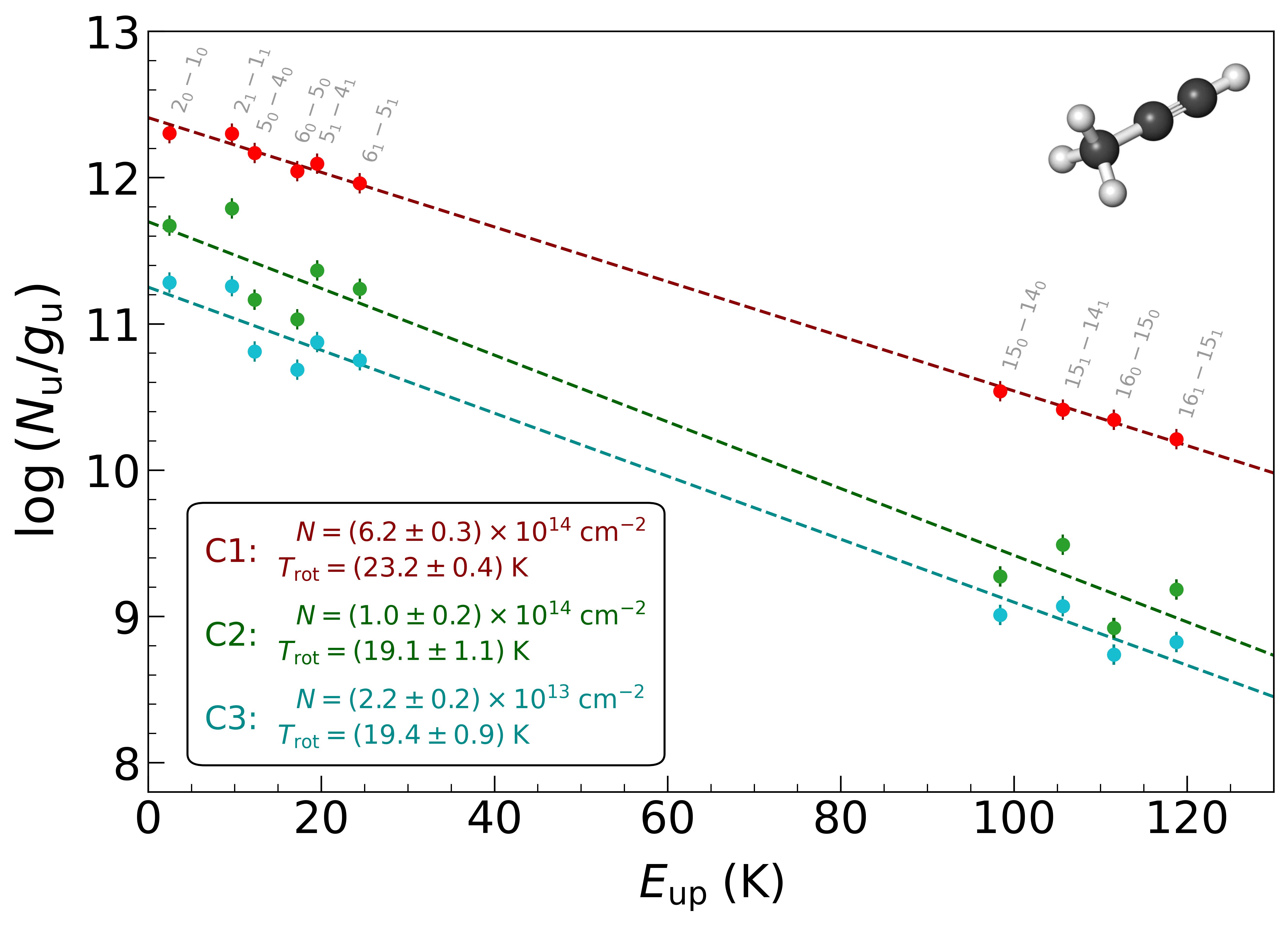}
    \end{subfigure}
    \caption{Rotational diagrams of \ce{CH3CN} (left) and \ce{CH3CCH} (right) computed using the $K=0,1$ components for each rotational transition ($J$) for all three velocity components characterised in G+0.633 (C1 in red, C2 in green and C3 in cyan colors). The dashed lines delineate the best linear fit to the data for each component, coloured matching its corresponding data points. The resulting column density ($N$) and rotational temperature ($T_{\text{rot}}$) estimates are accordingly presented framed within the black box. The specific transitions used (the same for all components) are displayed in grey and labelled using the notation for symmetric tops ($J_K$).
    }
    \label{fig:CH3CN_CH3CCH_RD}
\end{figure*}

Fig.~\ref{fig:CH3CN_CH3CCH_RD} shows the rotational diagram analyses of the three velocity components identified in G+0.633 for both \ce{CH3CN} and \ce{CH3CCH}, deriving the rotational temperatures ($T_{\text{rot}}$) and column densities ($N$). For each component, the analysis was based on the integrated intensities of the $K=0$ and $K=1$ transitions over the corresponding linewidths, accounting for the contributions of all molecules detected in the cloud as well as those from the other two velocity components. For \ce{CH3CN}, we obtained $T_{\text{rot}}$ of $13.7\pm0.6$\K, $12.4\pm1.5$\K and $12.0\pm1.2$\K, and $N$ of $(5.7 \pm 0.9)\times10^{13} \; \text{cm}^{-2}$, $(9.4 \pm 0.2)\times10^{12} \; \text{cm}^{-2}$ and $(5.45 \pm 0.35)\times10^{12} \; \text{cm}^{-2}$ for C1, C2 and C3, respectively. In the case of \ce{CH3CCH}, the derived $T_{\text{rot}}$ are $23.2\pm0.4$\K, $19.1\pm1.1$\K and $19.4\pm0.9$\K, and the resulting $N$ are $(6.2 \pm 0.3)\times10^{14} \; \text{cm}^{-2}$, $(1.0\pm0.2)\times10^{14} \; \text{cm}^{-2}$ and $(2.2 \pm 0.2)\times10^{13} \; \text{cm}^{-2}$ for C1, C2 and C3, respectively.

\section{Details of \ce{HC3N} non-LTE analysis used to constrain the gas volume density ($n_{\ce{H2}}$) in G+0.633}
\label{appendix:nonLTE_HC3N}

\begin{figure*}[ht!]
    \centering
    \includegraphics[width=\textwidth]{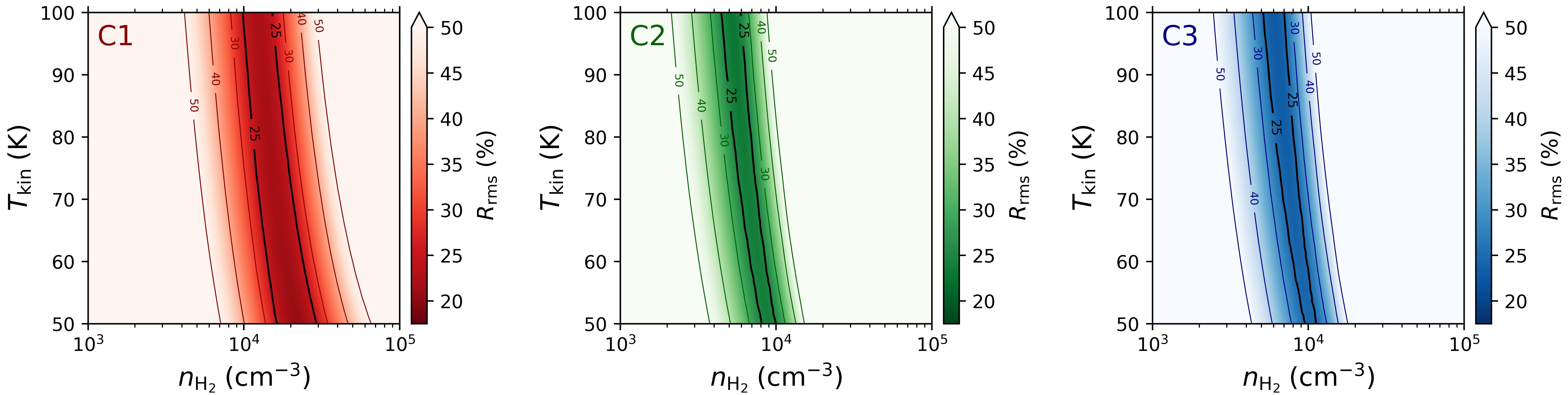}
    \caption{\textsc{RADEX} grids used to constrain $n_{\ce{H2}}$ in G+0.633 from the observed \ce{HC3N} emission. Colours represent the root-mean-square relative deviation ($R_{\rm rms}$, in \%) between the observed and predicted integrated intensities for each velocity component using the seven detected transitions. Black contours mark $R_{\rm rms} = 25\%$, defining the family of models that best reproduce the observations.
    } 
    \label{fig:HC3N_nonLTE_models_grid} 
\end{figure*}

In this appendix, we detail the methodology adopted to constrain $n_{\ce{H2}}$ in G+0.633 through \ce{HC3N} non-LTE analysis. To quantify the agreement between the observed \ce{HC3N} emission and the non-LTE predictions, we computed a fractional $\chi^2$ estimator ($\chi^2_{\rm frac}$) for each model in the \textsc{RADEX} grid. For a given model, the statistic was defined as $\chi^2_{\rm frac} \equiv \sum_{J} \left(\Delta I_J/I_{J, \, {\rm obs}}\right)^2$, where $\Delta I_J \equiv I_{J, \, {\rm mod}} - I_{J, \, {\rm obs}}$ is the difference between the integrated intensities predicted by the model ($I_{J, \, {\rm mod}}$) and observed ($I_{J, \, {\rm obs}}$) for the seven detected \ce{HC3N} transitions ($J$). These values were converted into a root-mean-square relative deviation, $R_{\rm rms} \equiv 100\times(\chi^2_{\rm frac}/7)^{1/2}$, which can be directly interpreted as the average percentage discrepancy between the predicted and observed integrated intensities. The resulting $R_{\rm rms}$ maps are shown in Fig.~\ref{fig:HC3N_nonLTE_models_grid}. To identify the family of models that satisfactorily reproduce the observations, we adopted a threshold of $R_{\rm rms}=25\%$, shown as black contours in Fig.~\ref{fig:HC3N_nonLTE_models_grid}. This elevated value reflects a systematic deviation driven by the $J=5-4$ transition, which affects only the global residual level and has no impact on the inferred $n_{\rm H_2}$. This criterion yields $n_{\ce{H2}}$ values of $\sim$(1.1$-$2.5)$\times10^4$, $\sim$(0.5$-$0.9)$\times10^4$ and $\sim$(0.6$-$1)$\times10^4 \; {\rm cm}^{-3}$ over $T_{\text{kin}} = 55-90$\K for C1, C2 and C3, respectively.

\clearpage

\section{\ce{HNCO} LTE analysis towards G+0.633}
\label{appendix:HNCO_LTE_analysis}

In this appendix, we provide the details of the LTE analysis we performed on \ce{HNCO} towards G+0.633, which exhibits emission in the three velocity components characterised in this cloud. For each of them, we conducted separate fits of the $K_{\text{a}} = 0$ and $K_{\text{a}} = 1$ ladders to mitigate as much as possible non-LTE effects, mimicking the same strategy used in G+0.693 \citep{Zeng2018} and therefore enabling the direct comparison between both clouds. To this aim, the original \ce{HNCO} spectroscopic entry was divided into three independent sub-entries within \textsc{MADCUBA} corresponding to the $K_{\text{a}}=0$, $K_{\text{a}}=1$ and $K_{\text{a}}>1$ ladders, with the partition function recomputed accordingly for each of them. We note that the $K_{\text{a}}>1$ entry was excluded in the analysis since no $K_{\text{a}}>1$ transitions are detected in the survey. We fitted the C1 component for the $K_{\text{a}} = 0,1$ lines by leaving all four physical parameters free, with their best estimates shown in Table~\ref{tab:physical_parameters_molecules_modelling}. By summing the column density estimates derived from the separate $K_{\text{a}}=0$ and $K_{\text{a}}=1$ fits, we derived a total \ce{HNCO} column density for C1 of $(3.16 \pm 0.17)\times10^{15}\, \text{cm}^{-2}$, which translates into a molecular abundance with respect to \ce{H2} of $(5.3 \pm 1.1)\times10^{-8}$. For the C2 and C3 components, we performed the analysis of their identified $K_{\text{a}} = 0$ and/or $K_{\text{a}}=1$ transitions (only detected for C2) by fixing the $\text{FWHM}$ to 13\kms and 18\kms, respectively. The remaining parameters were left free, deriving total \ce{HNCO} column densities and corresponding molecular abundances relative to \ce{H2} of $(4.7 \pm 0.3)\times10^{14}\, \text{cm}^{-2}$ and $(6.7 \pm 2.1)\times10^{-9}$ for C2 (summing the estimates from the separate $K_{\text{a}}=0$ and $K_{\text{a}}=1$ fits as for C1), while $(9.6 \pm 2.3)\times10^{13}\, \text{cm}^{-2}$ and $(3.4 \pm 1.2)\times10^{-9}$ for C3 (derived only from the detected $K_{\text{a}}=0$ lines), respectively. The results of the LTE analysis on these other two components are also given in Table~\ref{tab:physical_parameters_molecules_modelling}.

\begin{figure*}[ht!]
    \centering
    \includegraphics[width=\textwidth]{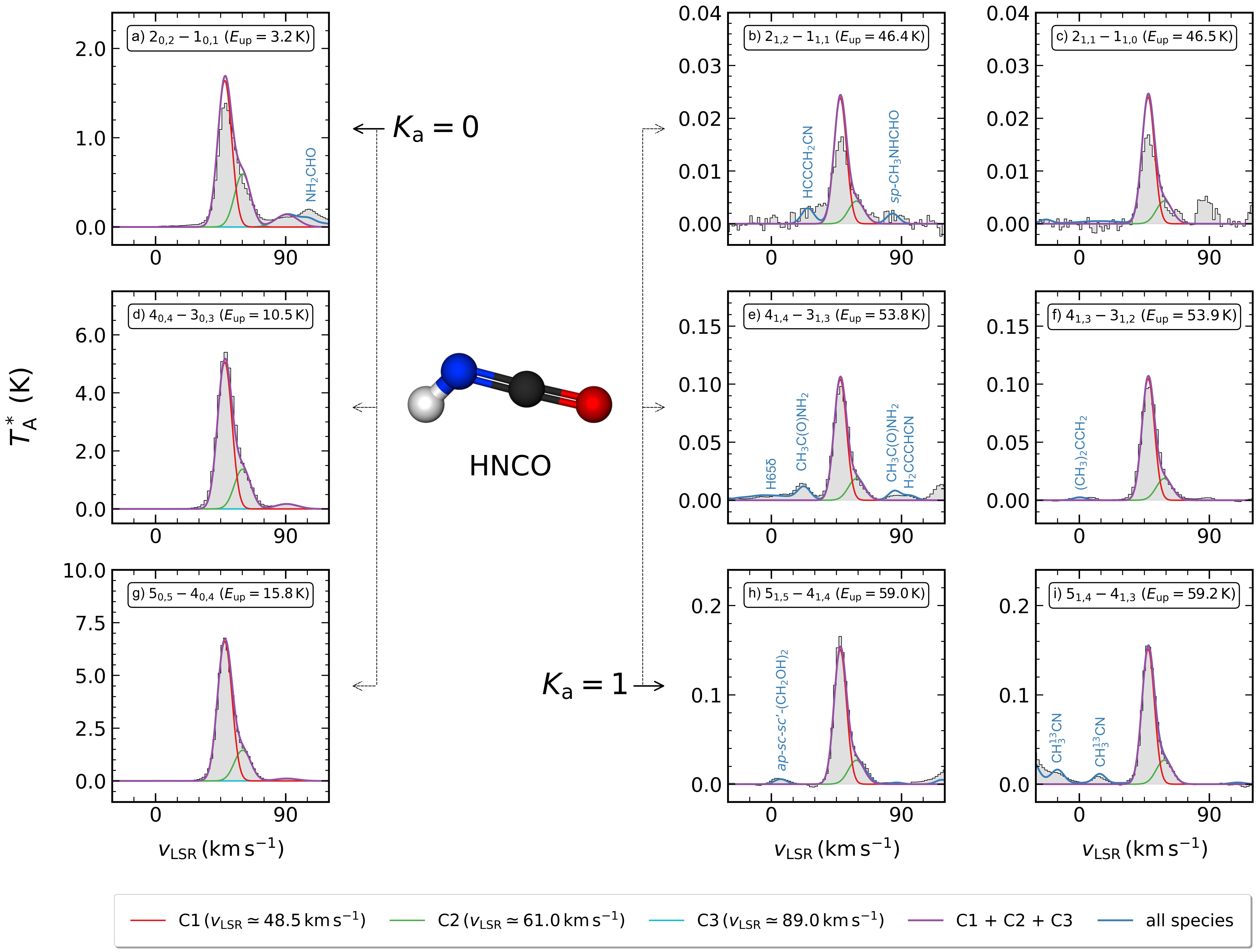}
    \caption{Same as Figs.~\ref{fig:CO_isotopologues} and \ref{fig:HC3N_low-J} (see main text), but for \ce{HNCO}. The least energetic $K_{\text{a}} = 0$ transitions detected in G+0.633 are displayed in the column at left, and their two corresponding $K_{\text{a}} = 1$ lines also traced are shown in the right columns. Panels labels also include the energy of the upper level involved in each transition. The molecular structure of \ce{HNCO} in shown in the centre (carbon atoms in grey, nitrogen atoms in blue, oxygen atoms in red and hydrogen atoms in white colours).} 
    \label{fig:HNCO} 
\end{figure*}

\clearpage

\section{Class I methanol masers and Radio Recombination Lines (RRLs) detected in G+0.633}

\begin{figure*}[ht!]
    \centering
    \includegraphics[width=\textwidth]{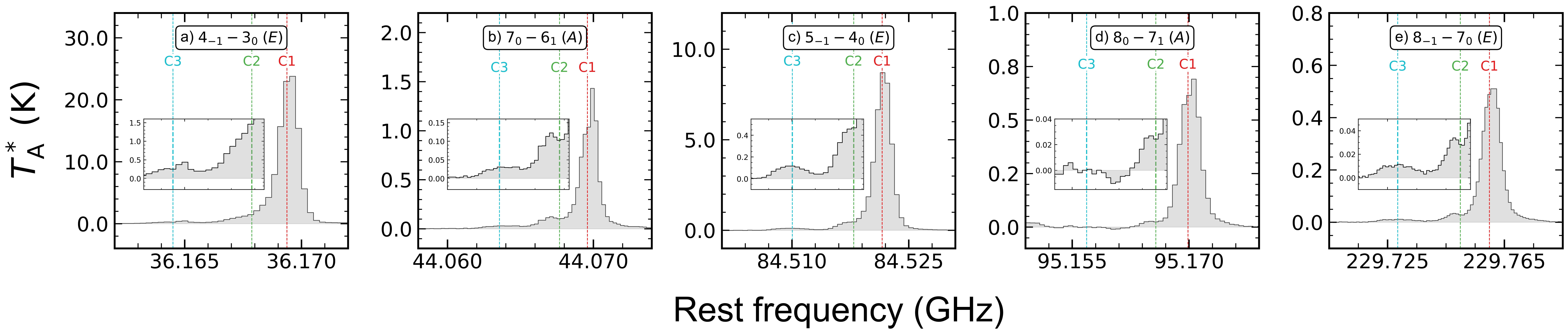}
    \caption{Spectra of the Class I methanol masers detected towards G+0.633. The dashed red, green and cyan vertical lines denote the central velocities of G+0.633 C1, C2 and C3 components, respectively. Panel labels indicate the particular transitions being plotted.} 
    \label{fig:CH3OH_ClassI_masers} 
\end{figure*}

Fig.~\ref{fig:CH3OH_ClassI_masers} displays the observational signatures of all Class I methanol masers identified in the current G+0.633 survey. All of them exhibit very bright emission at the velocities of the main C1 component (peaking at $\sim$48.5\kms), highly contrasting with the much fainter profiles from the C2 and C3 components (peaking at $\sim$61\kms and $\sim$89\kms, respectively). The linewidths of the C1 masers ($\sim$7\kms) are reasonably similar to those of G+0.633 lines ($\sim$10\kms) and identical for the other two components ($\sim$13\kms for C2, $\sim$18\kms for C3), indicating that the observed maser emission originates from the same gas reservoirs.

\begin{figure*}[ht!]
    \centering
    \includegraphics[width=\textwidth]{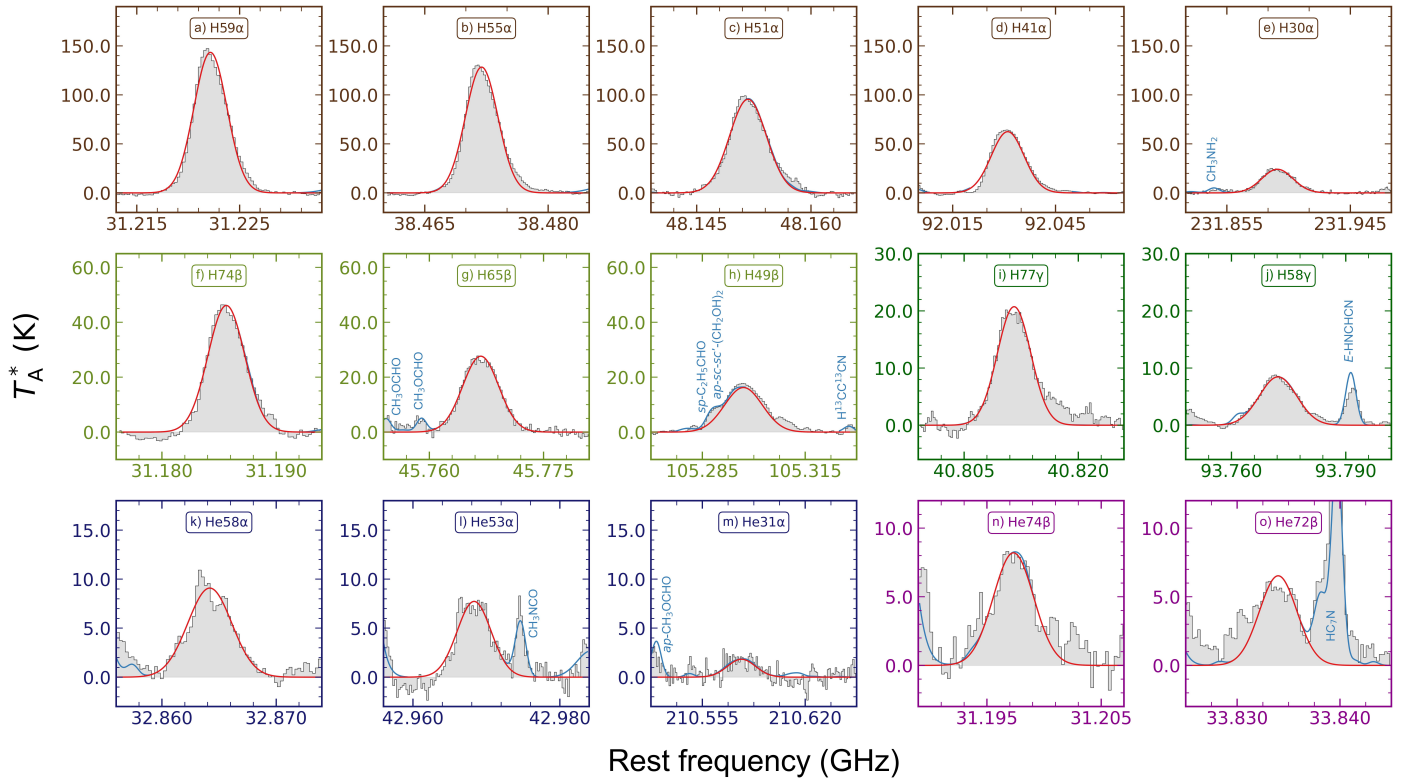}
    \caption{Selected unblended Hydrogen (top two rows) and Helium (bottom row) RRLs detected towards G+0.633, with velocities $\sim$55$-$65\kms and $\text{FWHM}$ of $\sim$30$-$40\kms. Overlaid to the observed spectrum (gray-shaded), the spectral model of each RRL and the contribution from all species identified in the cloud is shown in red and blue, respectively. Panels displaying transitions of the same category ($\alpha$, $\beta$, or $\gamma$) for either \ce{H} or \ce{He} RRLs (indicated by the panel labels) are outlined with matching boundary colours.} 
    \label{fig:RRLs_G0633} 
\end{figure*}

Another distinctive feature of the G+0.633 molecular cloud is the detection of numerous very energetic Hydrogen and Helium Radio Recombination Lines (RRLs), indicative of heavily ionised gas in the line of sight. This emission is found to originate locally from Sgr B2(DS), previously interpreted as an expanding HII region \citep{Meng2019}. It appears unrelated to any of the three velocity components identified in G+0.633, since the RLLs exhibit peak velocities of $\sim$55$-$65\kms and very broad linewidths of $\sim$30$-$40\kms. Fig.~\ref{fig:RRLs_G0633} displays a representative subset of the detected RRLs, selected among the $\sim$100 identified across the survey. All these span a wide range of energy levels ($n_\text{up}$ down to 29) and include large quantum leaps ($\Delta n \leq 8$, $\alpha$ to $\theta$-transitions, for Hydrogen RRLs; and $\Delta n \leq 2$, $\alpha$ and $\beta$-transitions, for \ce{He} ones). Each RRL found in the G+0.633 survey was fitted independently, adopting a $T_\text{ex}$ of $\sim$8000\K (as in \citealt{Meng2019}) and leaving the remaining parameters free. This approach stems from RRLs non-LTE excitation already reported by \citet{Meng2019}, manifested in a peculiar intensity–frequency anti-correlation (Fig.~\ref{fig:RRLs_G0633}).

\end{appendix}

\end{document}